\pdfoutput=1
\documentclass[smallextended]{svjour3}       
\smartqed  
\usepackage{graphicx} 
\usepackage[font=small,labelfont=bf]{caption}
\usepackage{wrapfig}
\usepackage{lipsum}
\usepackage{graphicx}
\usepackage{adjustbox}
\usepackage{footnote}
\makesavenoteenv{tabular}
\makesavenoteenv{table}
\usepackage{array}
\usepackage{amsmath}
\usepackage{algorithm}
\usepackage{subcaption}
\usepackage{hyperref}
\usepackage[noend]{algpseudocode}
\usepackage[parfill]{parskip}
\makeatletter
\def\BState{\State\hskip-\ALG@thistlm}
\makeatother
\algnewcommand{\algorithmicand}{\textbf{ and }}
\algnewcommand{\algorithmicor}{\textbf{ or }}
\algnewcommand{\OR}{\algorithmicor}
\algnewcommand{\AND}{\algorithmicand}

\begin{document}

\title{Information Extraction Framework to Build Legislation Network
}
\subtitle{}


\author{Neda Sakhaee \and 
Mark C Wilson 
}


\institute{N Sakhaee \at
              Computer Science Department, University of Auckland\\
              \email{nsak206@aucklanduni.ac.nz}           
           \and
           M C Wilson \at
              Computer Science Department, University of Auckland\\
              \email{mc.wilson@auckland.ac.nz}
}

\date{Received: date / Accepted: date}

\maketitle

\begin{abstract}
This paper concerns an Information Extraction process for building a dynamic Legislation Network from legal documents. Unlike supervised learning approaches which require additional calculations, the idea here is to apply Information Extraction methodologies by identifying distinct expressions in legal text and extract quality network information. The study highlights the importance of data accuracy in network analysis and improves approximate string matching techniques for producing reliable network data-sets with more than 98 percent precision and recall. The values, applications, and the complexity of the created dynamic Legislation Network are also discussed and challenged.
\end{abstract}

\section{Introduction}
\label{intro}
\textbf{Legislation Networks} were first introduced in 2015 \cite{MKoniaris2017} and further discussed in \cite{NSakhaee2016} \cite{NSakhaee2017}. These networks are essential to explore the relationship between legislation and societies' evolution \cite{NSakhaee2016}. There are many obvious benefits from studying Legislation Networks \cite {NSakhaee2016} \cite{NSakhaee2017} \cite{PZhang2007} \cite{Fowler2007}, but building these networks is not always a straightforward task, as only a few legislation systems provide machine-readable documents or structured databases \cite {EURlex} \cite{Metalex}. The majority of legislation systems supply legal documents in human-readable format. For example the New Zealand Parliamentary Council Office provides machine-readable XML files \cite{PCO} only for the current active Acts, which constitute around ten percent of the entire set of Acts. All other historic documents are scanned and supplied by a third party institute in Portable Document Format (PDF) \cite{NZLII}. 


To extract information from legal documents, the first step is the conversion of images into text if the text is not available. This concept is well studied as \textbf{optical character recognition (OCR)} \cite{OCR}, and there are several techniques and tools developed to convert typewritten or handwritten images to text. OCR is the first step of the proposed framework, and for the case study we selected ABBYY FineReader \cite{ABBYY}.

 \textbf{Information Extraction (IE)}  involves locating and extracting specific information from text \cite{IE0}. Information Extraction assumes that in each individual text file there are one or more entities that are similar to those in other text documents but differing in the details \cite{IE2}. 

IE approaches in the legal domain are considerably different from other knowledge areas because of the two main characteristics of legal texts. Legal documents exhibit a wide range of internal structure, and they often have a significant amount of manual editorial value added.
One of the earliest information retrieval approaches for legal materials based on searching inside the document was proposed in 1978 \cite{IE1}. Later works mainly used natural supervised learning techniques to retrieve the required data from legal texts, but with a substantial error \cite{Cheng2009} \cite{Textnailing2017}. In the proposed framework of this study several IE tasks are used, and more are described later in this section.


\textbf{Named entity recognition (NER)}
is one of the main sub-tasks of IE. The goal of this task is to find each occurrence of a named entity in the text \cite{NER1}. Entities usually include people, locations, quantities, and organizations but also more specific entities such as the names of genes and proteins \cite{NER2}, the names of college courses \cite{NER3}, and drugs \cite{NER4}. In the New Zealand legislation corpus, entities could be the name of legislative documents such as Acts, Regulations, Bills, Orders, or Case-Laws \cite{NSakhaee2016}. In the case study which is discussed \autoref{Application}, the main required entities inside the text documents are is the names of the New Zealand Acts. 

The main traditional NER algorithm that identifies and classifies the named entities is statistical sequence modeling \cite{NER1}. But there are other modern approaches based on combinations of lists, rules, and supervised machine learning \cite{NER5}. To extract the require information for Legislation Network, there are clear rules to identify the named entities, and the classification of the entities is not needed. Therefore the second NER approach is more appropriate and discussed further for the proposed framework.

The next IE task which is used in our study is to detect the relationships that exist among the recognized entities. This task is called \textbf{relation extraction (RE)} \cite{NER1}.
The earliest algorithm for relation extraction is the use of lexico-syntactic patterns \cite{RE1}. This algorithm is still valid and widely used, but there are other algorithms introduced later such as supervised learning \cite{NER1} and bootstrapping \cite{boot}\cite{boot1}.
Considering that legislation texts are well structured, it is assumed that there is a large collection of previously annotated material that can define the rules for classifiers.

\textbf{Approximate string matching} techniques find items in a database when there may be a spelling mistake or other error in the keyword \cite{ASM1}. This is becoming a more relevant issue for fast-growing knowledge areas such as information retrieval \cite{ASM4}. Various techniques are studied to address the identity uncertainty of the objects, and briefly are reviewed in this study. These techniques could be distance based, token based, or a hybrid model of the distance and token based models.\\~
Damerau-Levenshtein metrics are the main approximate string matching techniques to address the distance functions between the two strings \cite{ASM2} \cite{ASM3}. The most famous function in this category is \textit{edit-distance}, and it is defined as the minimum number of changes required to convert one string into the other \cite{ASM3}. Several alternative distance functions to the edit-distance have been proposed such as $q$-gram and maximal matches \cite{ASM5}.\\~
The next set of techniques are token based or probabilistic object identification methods adapted for string matching tasks \cite{FSM1}\cite{FSM2} \cite{ASM1}. \textit{Jaccard similarity} and \textit{cosine similarity} are common token based measures widely used in the information retrieval community\cite{FSM1}. Hybrid techniques combine distance-based and token-based string matching measures such as Jaro-Winkler \cite{FSM4}. All of the string matching algorithms have been developed by filtering and bit-parallelism approaches.\\~
The fastest algorithms use a  combination of filters to discard most of the text by focusing on the potential matches. Hybrid models significantly improve precision and recall reducing the error in a range between $0.1$ to $0.2$ \cite{ASM4}.

Network inferences require high accuracy of data \cite{Error1}.  For this study various string matching techniques are examined for the Legislation Network and comparing the results, a hybrid model of \textit{Jaccard similarity} and \textit{edit-distance} is used as described in the next section.

\textbf{The main contribution of this study} is the proposed Information Extraction framework which engages several processes and enables the researcher to have access to the network information from historic documents. This framework makes it possible to study the Legislation Network as a dynamic graph. In this paper the case study covers all Acts in New Zealand legislation corpus including historic, expired, repealed and consolidated Acts as at end of September 2018. This comes to a set of 23870 PDF files of which about $87\%$ are in scanned image format. 

\begin{figure}
    \centering
    \includegraphics[scale=.7]{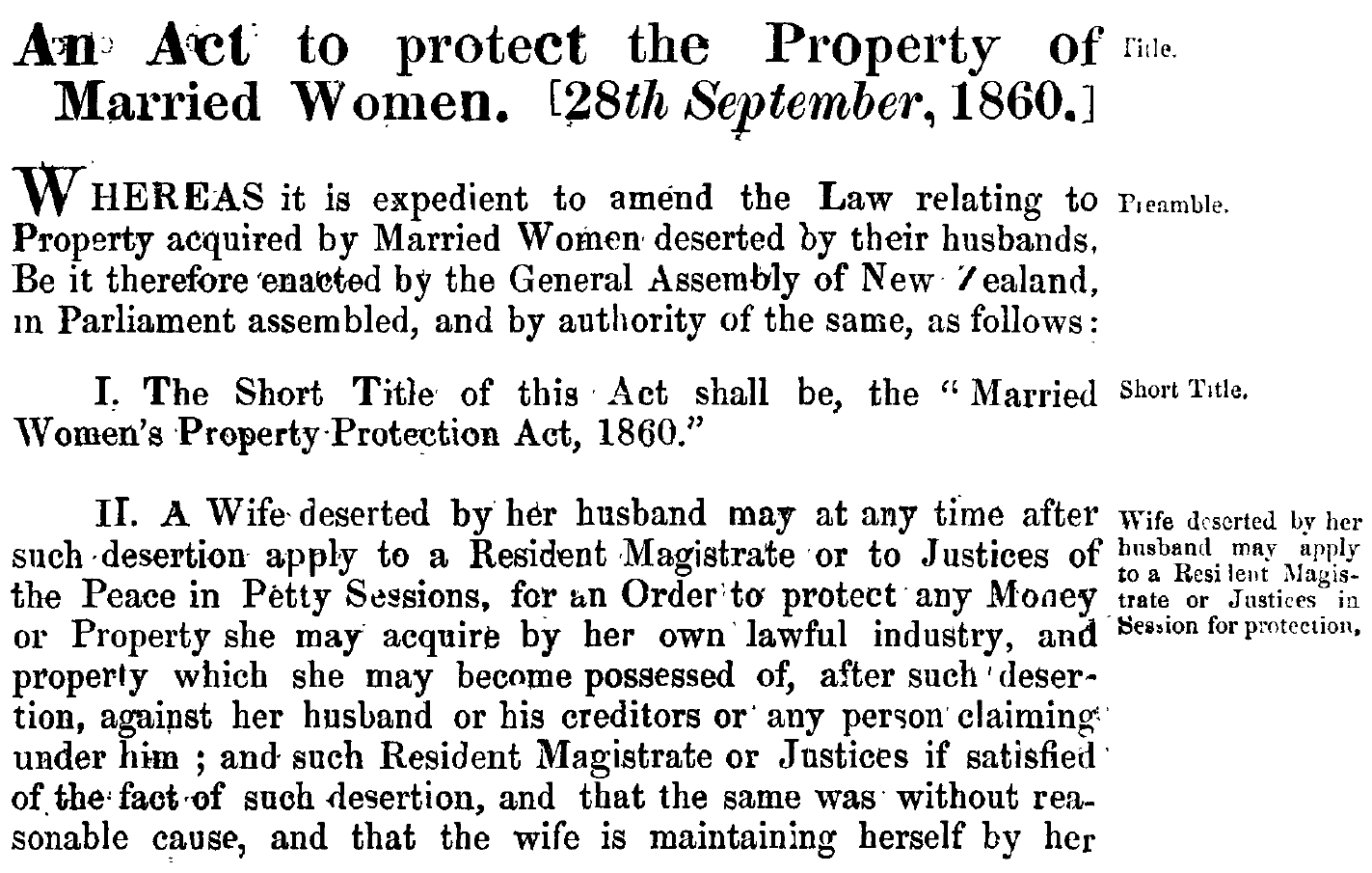}
    \caption{Married Women Property Protection Act 1860}
    \label{sample}
\end{figure}

\autoref{sample} shows a sample image of an average quality scanned PDF document. The proposed framework suggests a high-performance procedure to derive network information from such poor quality documents. In the following sections, examples and the experimental results are used to illustrate the framework, its performance, and its potential applications. 

In this section a summary of the required Information Extraction processes and methodologies is discussed. In the next section the proposed framework is presented and various examples are explained. Then the case study analysis and the application of the proposed framework are examined. Next, a number of experiments are designed and studied to evaluate the accuracy of the extracted information and to study the robustness of Legislation Network. The study finishes with a quick review on the novelty and the importance of discovering the time-varying behaviour of the Legislation Network.
\section{The proposed Information Extraction framework}
\begin{wrapfigure}{l}{5.5cm}
\includegraphics[scale=.7]{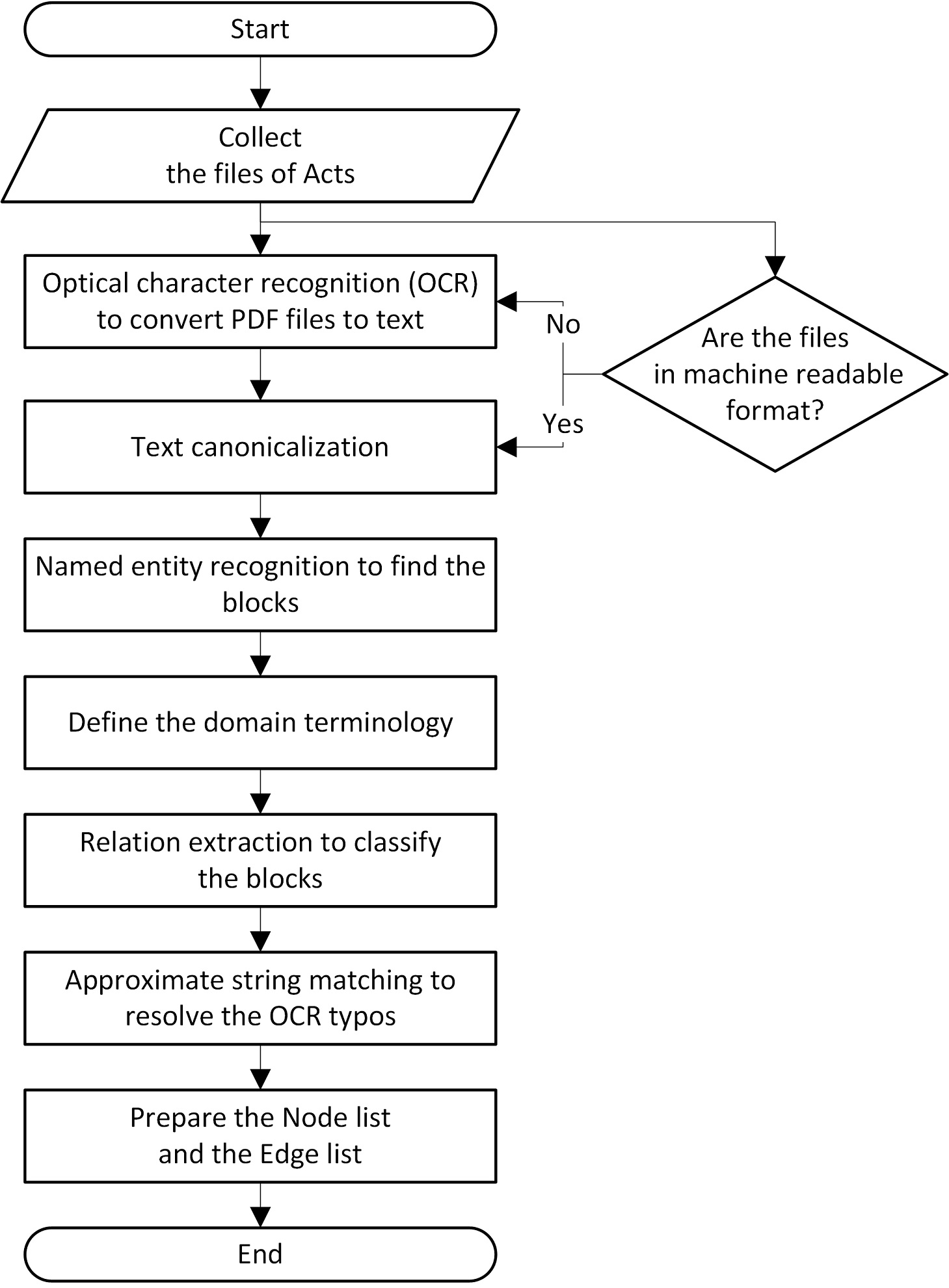}
\caption{Legal Text IE Framework}
\label{IEF}      
\end{wrapfigure}

In this section, the Information Extraction framework to build the Legislation Network is discussed. \autoref{IEF} depicts the overview of the proposed framework.\\~
The process starts with the conversion of non machine readable files to text by using \textit{OCR} available tools. This step is relatively straightforward, but could be time-consuming considering the number of documents in the study. As mentioned earlier, in the case study the tool named ABBYY FineReader \cite{ABBYY} is used. The average accuracy of this step is just above 80 percent and implies the need for a typos analysis step that is discussed in section \ref{Approximate String Matching}.
\subsection{Text Canonicalization}
The next step in the proposed framework is \textit{text canonicalization} \cite{canon}. There are several required tasks to convert all of the text files into a unique format, so the rules can be defined more easily while running the Information Extraction tasks. The text canonicalization step could be implemented via different approaches depending to the text style and language. In this paper, some of the common tasks are suggested, and two potentially required tasks are described.\\~
In the case study the designed system transfers all letters to \textit{Lowercase}. This transition applies a level of consistency across the text documents and the Information Extraction rules.
In the experiments, the system also replaces \textit{Special Characters} with generic tags in the text. The only character which is not replaced is the parenthesis, as it is often used in the title of legislation.
The other suggested generic text canonicalization task is to replace multiple spaces with one space.

Apart from the general text canonicalization steps, there are other potential corrections that shape the text to make it a better input for the Information Extraction process. The first is to remove text margins that OCR mistakenly merges to the main body of text. \autoref{sample} includes examples of these margins that might impact the Information Extraction rules and result in error. As an example in the case study, the phrase \textit{short title} is often used in the margin, and OCR merges it to the nearest part of the text. This might impact the named entity recognition task, so the system removes this phrase and many of its possible misspelled forms from the text. 
The next recommended text canonicalization step is to resolve the misspelling issues for the keywords used in the Information Extraction process. As an example in this study, Acts are the main entities, and the rule to recognize them uses the keyword \textit{Act}. So the system corrects some of the possible misspelling forms of the word \textit{-act-}.
\begin{table}[h]
    \centering{}\protect\caption{OCR and Text canonicalization result comparison}
    \label{tab:canonicalization}
    \adjustbox{max height=\dimexpr\textheight-6.5cm\relax,
           max width=\textwidth}{
    \begin{tabular}{l l l}
    \hline
   OCR & I$\sim$ The Short Title' of this' Act shall be, the ((Married Short TItle,'Vomen's 'Property'Protectio\pounds  Aot, 1860." \\
   \hline
   Text canonicalization  & I the short title of this act shall be the ((married vomens propertyprotectio act 1860\\
   \hline
    \end{tabular}
}
\end{table}
To provide a better explanation of this step, \autoref{tab:canonicalization} provides an example referring to third paragraph of the \autoref{sample} image. 
As can be seen the text canonicalization converts the text to a simpler unique structure prepares it for the next steps of \textit{named entity recognition} and \textit{relation extraction}. Typo resolution is not expected at this step, being covered under the last step via Approximate String Matching.
\subsection{Named Entity Recognition and Relation Extraction}
As explained, the text canonicalization step normalizes the text files to a unique format and prepares them for in-depth information extraction steps. To extract the network node information, a combined Named Entity Recognition approach is suggested which engages rules and supervised learning. To identify the rules, a sample set of the documents should be reviewed. The sample size is not necessarily large, but a stratified sampling approach is suggested to eliminate the impact of time-period style and author's writing style. 

\autoref{Entities} shows examples of the entities in the case study based on the Acts recognition rules. Acts are the main part of the New Zealand legislation system and as explained before the case study only considers the Acts. 
\begin{table}[h]
\centering{}\protect\caption{Entities, types and examples}
\label{Entities}
\adjustbox{max height=\dimexpr\textheight-6.5cm\relax,
           max width=\textwidth}{
\begin{tabular}{ l  l  l  l }
\hline
Type & Tag & Sample & Canonicalized text \\ 
 \hline
 Year & YR & 1860 & the short title of this act shall be the ((married vomens propertyprotectio act 1860 \\ 
 Act & ACT & married vomens propertyprotectio act 1860 &  the short title of this act shall be the ((married vomens propertyprotectio act 1860\\ 
 \hline	
\end{tabular}
}
\end{table}
In the case study stratified sampling method is used and the strata are five different time periods in a range of more than 200 years. A total number of 55 text files are reviewed, and several clear rules engaging a set of keywords and lists are built to identify the named entities. \autoref{NERrules} provides examples of the rules in each stratum and $y$ represents the year in which the Act is commenced. 

\begin{table}[h]
\centering{}\protect\caption{Examples of the Named Entity Recognition rules}
\label{NERrules}
\adjustbox{max height=\dimexpr\textheight-6.5cm\relax,
           max width=\textwidth}{
\begin{tabular}{ l  l  l  l }
\hline
Stratum & Keyword example & Rule example & Sample document  \\ 
 \hline
 $y<1850$ & \textit{ordinance} & an [\textit{keyword}] to \textit{any phrase} of [\textit{act name}] [date] & Police Magistrates Act 1841 \\
 $1850<y<1900$ & \textit{short title}, \textit{shall be} & the [\textit{keyword}] of this act [\textit{keyword}] the [\textit{act name}] [\textit{year}] & Customs Tariff Act 1873  \\ 
 $1900<y<1950$ & \textit{amend}, \textit{consolidate} & an act to [\textit{keyword}] \textit{any phrase} of the [\textit{act name}] [date] & Mining Act 1926 \\
 $1950<y<2000$ & \textit{meaning}, \textit{section} & same [\textit{keyword}] as in [\textit{keyword}] [any number] of the [\textit{act name}] [\textit{year}] & Copyright Act 1962 \\
 $2000<y$ & act & this [\textit{keyword}] is the [\textit{act name}] [\textit{year}] & Social Security Act 2018\\
 \hline	
\end{tabular}
}
\end{table}

Alongside recognizing the entities, to extract the network edge information, a rule based Relation Extraction approach is suggested considering that legislation texts are contextually structured. To identify the rules, this study suggests to use the same sample set which is used for the Named Entity Recognition. From the case study it is observed that the style of writing legislation has changed considerably over time, so the sampling approach is very important to minimize the impact of various text styles. By reviewing the sample files, there is a large collection of previously annotated material that can define the rules for relation classifiers. 

For the case study, as explained total number of 55 text files are reviewed, and several classifier rules engaging a set of keywords are built to identify the relations between the named entities. \autoref{Relations} summarizes the entity relation list for the case study and provides examples. 
\begin{table}[h]
\centering{}\protect\caption{Relations example}
\label{Relations}
\adjustbox{max height=\dimexpr\textheight-6.5cm\relax,
           max width=\textwidth}{
\begin{tabular}{ l  l  l  l }
\hline
Relation & Type & Canonicalized text & Sample document \\ 
 \hline
 Title & TIT & the short title of this act shall be the ((married vomen propertyprotectio act 1860 & Married Women Property Protection Act 1860 \\ 
 Citation & CIT & within the meaning of section 5 of the companies act 1993 & Trade Marks Act 2002 \\ 
 Amendment & AMD & section 25.1b amended, by section 5.2 of the trade marks amendment act 2005 & Trade Marks Act 2002\\ 
 Partial Repeal & PRP & section 5(1) repealed, by section 4(8) of the trade marks amendment act 2011 & Trade Marks Act 2002\\
 Repeal & FRP & acts repealed.
1860, No. 9.the married vyomens pfoperty protection act, 1860. & Married Women Property Protection Act 1880\\
  \hline	
\end{tabular}
}
\end{table}
This suggested process can be generalized for any other case study in Legislation Network building process considering that legislation texts are coherently structured. So there is always a large collection of previously annotated material that can define the rules for entity recognition and relation classifiers.
\subsection{Approximate String Matching}\label{Approximate String Matching}
Named entity recognition identifies the Acts and relation extraction recognizes the relationship between them. So these two steps result in an initial version of the node list and the edge list of the intended Legislation Network. However testing this network shows that the extracted data is shoddy with an average error rate of 12 \footnote{To  estimate  this error rate, a  cluster  sampling  method is  used  to  randomly  choose  ten  sets  of  30  entities.  By  manual  check  of  the samples, the rate of incorrectly matched entities is observed.} percent, so another step is required to resolve typo issues and imperfect entities. This poor-quality data implies the need for the approximate string matching step. To run this step two main components are required, the technique and the correct pattern. \autoref{ASMexample} provides an example which shows the first match as the output of the proposed approximate string matching technique. 
\begin{table}[h]
\centering{}\protect\caption{Approximate string matching example}
\label{ASMexample}
\adjustbox{max height=\dimexpr\textheight-6.5cm\relax,
           max width=\textwidth}{
\begin{tabular}{ l  l  l }
\hline
Technique & Extracted entity & First match \\ 
 \hline
Hybrid Model & married vomens propertyprotectio act 1860 & married women property protection act 1860 \\ 
 \hline	
\end{tabular}
}
\end{table}

As mentioned earlier after the implementation of different approximate string matching techniques, a hybrid model of \textit{Jaccard} and \textit{Edit-Distance} is designed and proposed.
\begin{figure}[h]
\centering
\includegraphics[scale=.5]{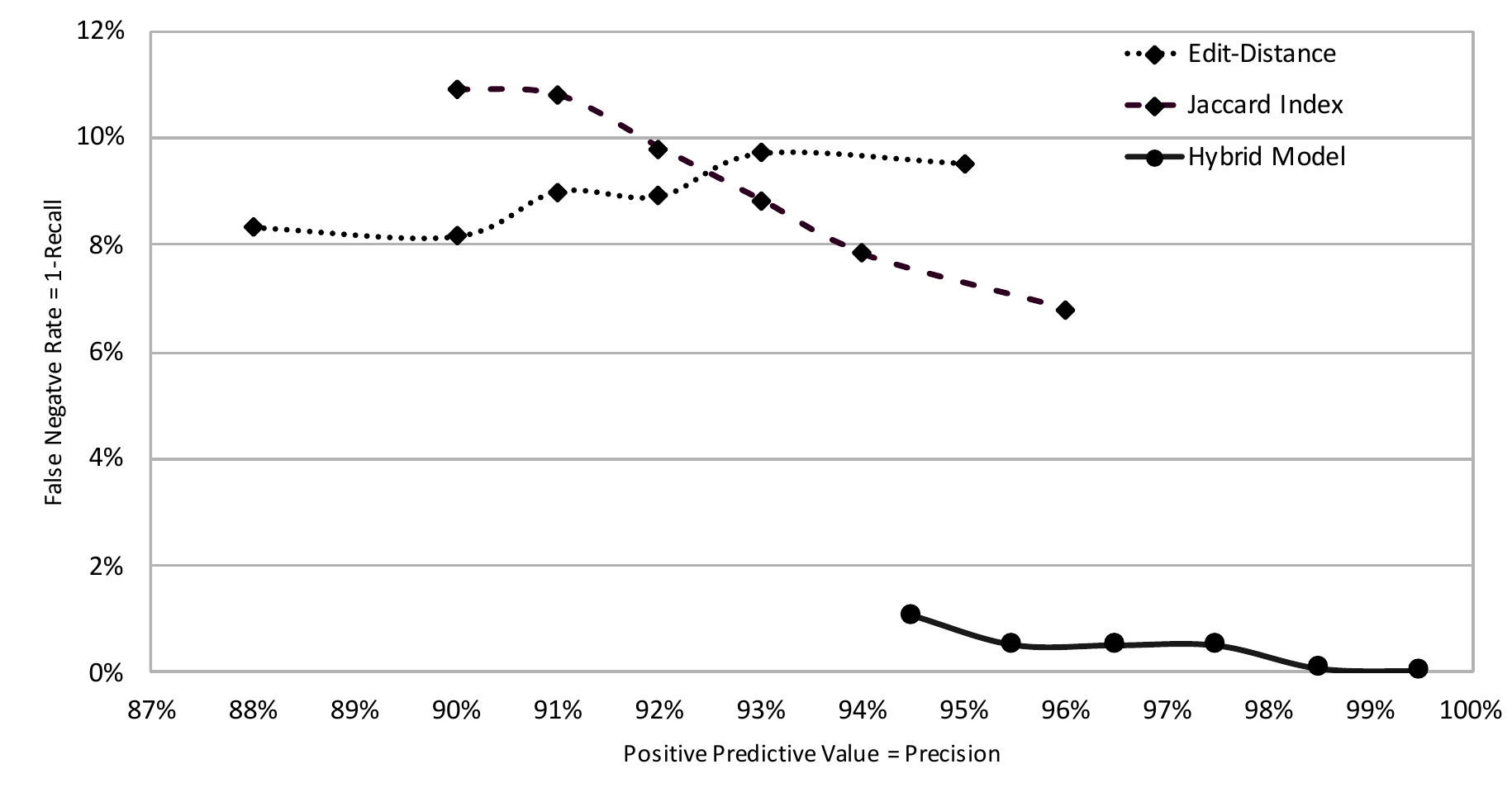}
\caption{Precision and Recall comparison of the approximate string matching techniques}
\label{ASMPic}
\end{figure}
Algorithm \ref{ASM} shows the proposed hybrid model, and \autoref{ASMPic} compares the results of the hybrid model with Edit-Distance and Jaccard techniques in terms of precision and recall of the approximate string matching step. To run this comparison, a stratified sampling technique is used with different time periods being the groups\footnote{Time periods: before 1800, 1800-1850, 1850-1900, 1900-1950, 1950-2000, 2000-2018}.

\begin{algorithm}
\caption{Approximate String Matching of Legislation}\label{ASM}
\begin{algorithmic}[1]
\Procedure{Legislation Name Matching}{}
\State $string1 \gets \textit{Extracted legislation name}$
\State $\textit{masterlist} \gets \text{Open }\textit{Legislation Title Master List}$
\State $j \gets \textit{1}$
\State $\textit{tline} \gets \text{The first line of }\textit{masterlist}$.
\State $GetOut \gets \textit{0}$
\While {$<GetOut \neq 0> \AND <tline \neq 0>$} \\~
\State $string2 \gets \textit{tline}$.
\State $m(j) \gets \textit{Jaccard(string1, string2)}$
\State $n(j) \gets \textit{EditDistance(string1, string2)}$
\If {$<m(j) = 0.5> \OR <n(j) = 0> $}
\State $\textit{GetOut}$
\EndIf
\State $j \gets j+1$. 
\State $\textit{tline} \gets \text{The next line of }\textit{masterlist}$
\State \textbf{close};
\EndWhile
\State $[x1 , I1] \gets \textit{max(m)}$
\State $[y1 , I2] \gets \textit{min(n)}$
\If {$y1 \textit{ is smaller than or equal to 5}$}
\State $ match \gets I2$
\ElsIf {$x1 \textit{ is bigger than 0}$}
\State  $ match \gets I1$
\EndIf
\EndProcedure
\end{algorithmic}
\end{algorithm}

The graph shows the error rates in each time sample of documents based on the chosen approximate string matching model. For example for the documents commenced prior to 1850, the first marker point at each graph line shows the false negative error and the precision of the chosen approximate string matching method.  
As can be seen samples from this oldest groups of acts show a higher error rate regardless of the approximate string matching method, and Edit-Distance performs slightly better for the old documents comparing to the Jaccard index. In summary the proposed hybrid model performs significantly better than the other two methods for all documents regardless of their age with less than two percent of false-negative error and average precision of more than 98 percent.

In the case study the pattern  which is used for the approximate string matching step is the list of all NZ Acts provided by NZLII \cite{NZLII}. In case of not having access to such a master list, the typo resolution could be more time consuming. Approximate string matching considerably improves the quality of the extracted information, result in reliable edge list and node list. Later in this study the evaluation of the final extracted data set and the robustness of the network is discussed. The robustness study proves the value of a high performing approximate string matching technique which improves the data quality significantly. 
\section{Application}\label{Application}
The proposed Information Extraction framework resolves the historic data limitation in previous studies \cite{NSakhaee2016}\cite{NSakhaee2017} and results in a large and reliable dynamic network data set which is called LegiNet and is available at \cite{dataverse}. This dynamic \cite{DynamicGraph} and complex network has a very intersecting range of characteristics and behaviours. To maintain the subject consistency of this paper, more in-depth analysis of network behaviours are delayed to the future studies. In this section, generic network science characteristics of the case studied network are discussed, and an overall view of the structural and node importance evolution is presented. 

\begin{figure}
\begin{center}
\begin{minipage}{0.3\linewidth}
\includegraphics[width=0.82\linewidth]{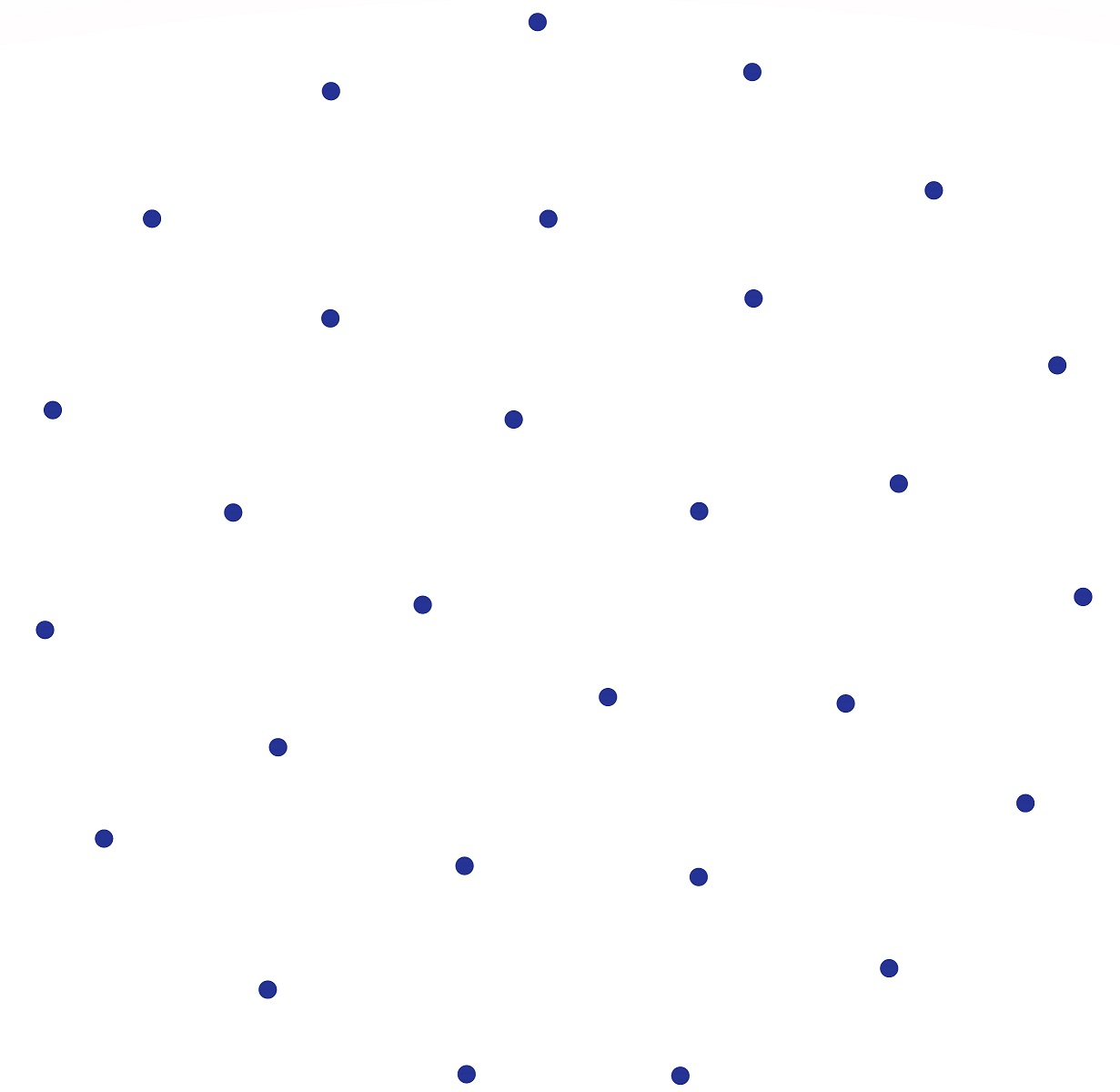}
\subcaption{1267-1839}\label{a1}
\end{minipage}%
\begin{minipage}{0.3\linewidth}
\includegraphics[width=0.82\linewidth]{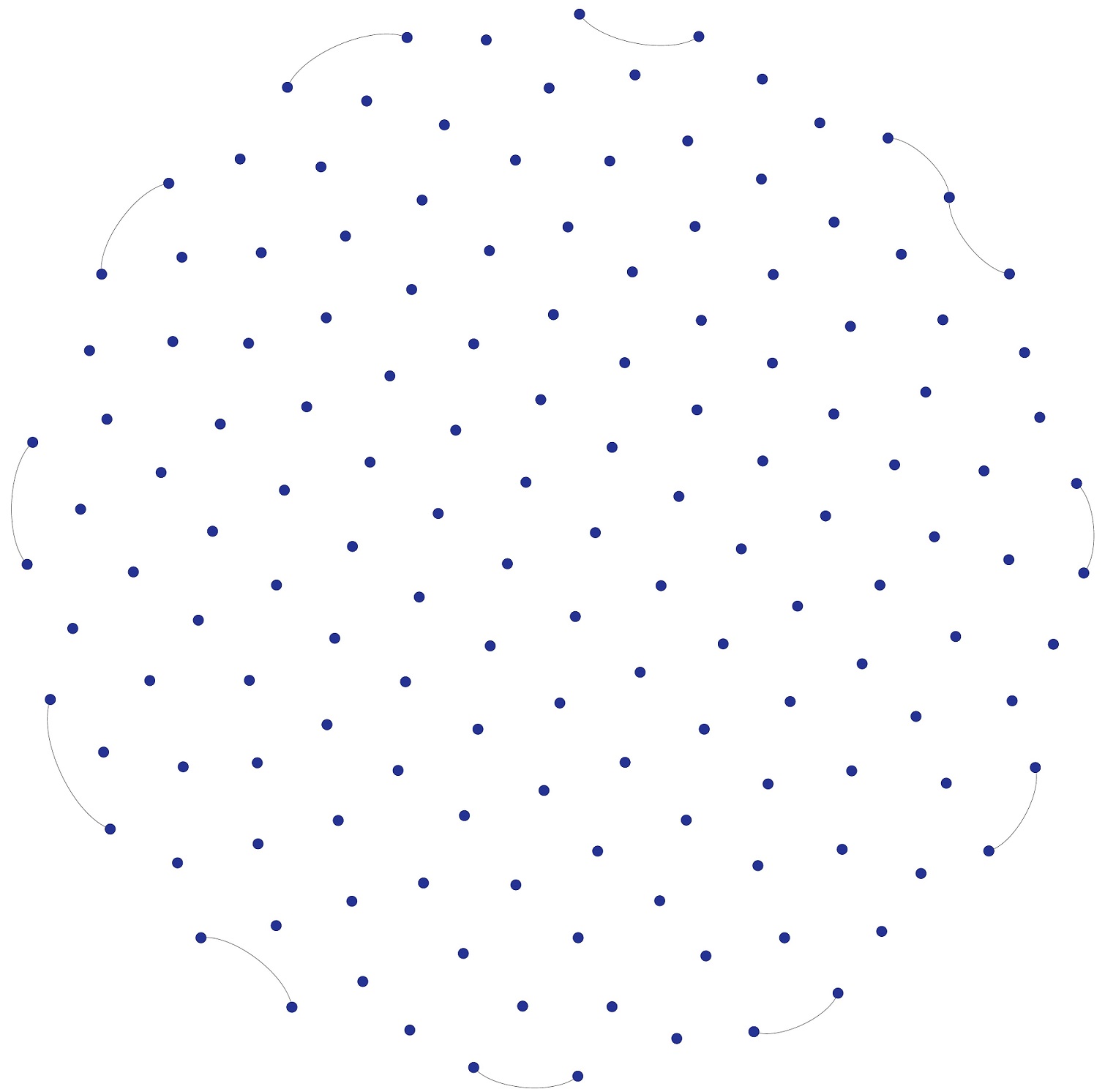}
\subcaption{1840s}
\end{minipage}%
\begin{minipage}{0.3\linewidth}
\includegraphics[width=0.82\linewidth]{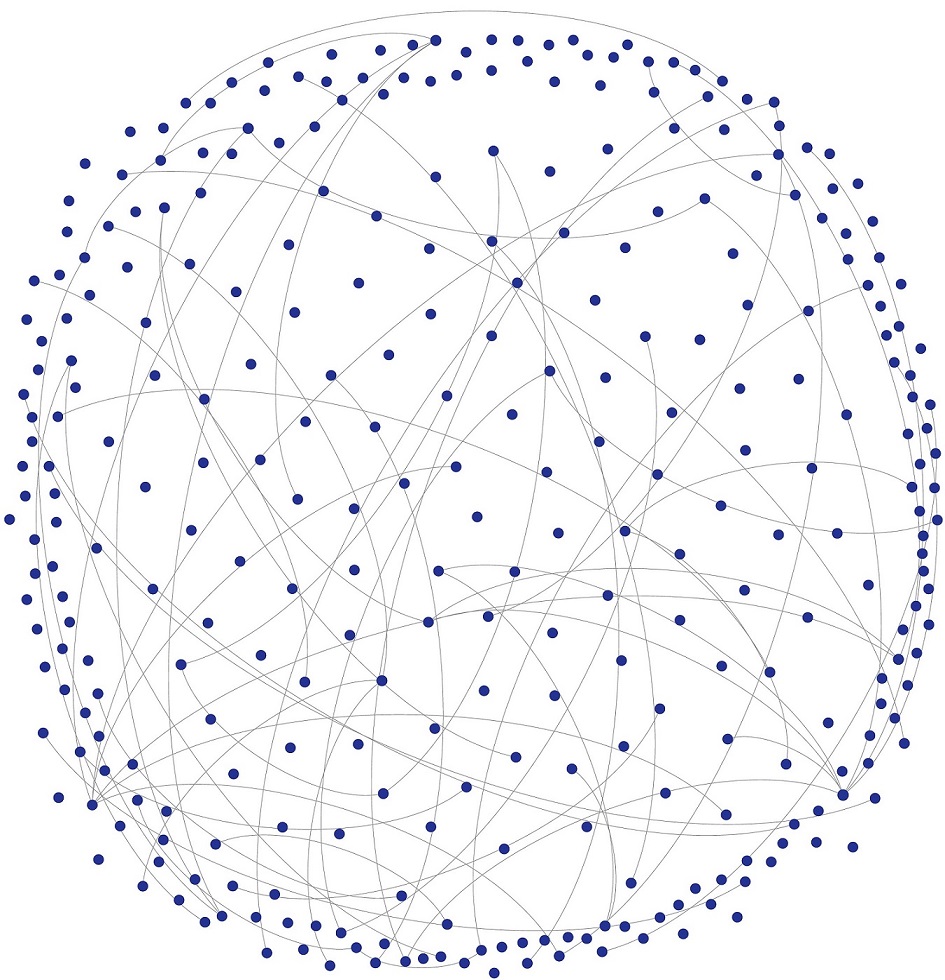}
\subcaption{1850s}\label{c1}
\end{minipage}%
\vspace{-\parskip}
\begin{minipage}{0.3\linewidth}
\includegraphics[width=0.82\linewidth]{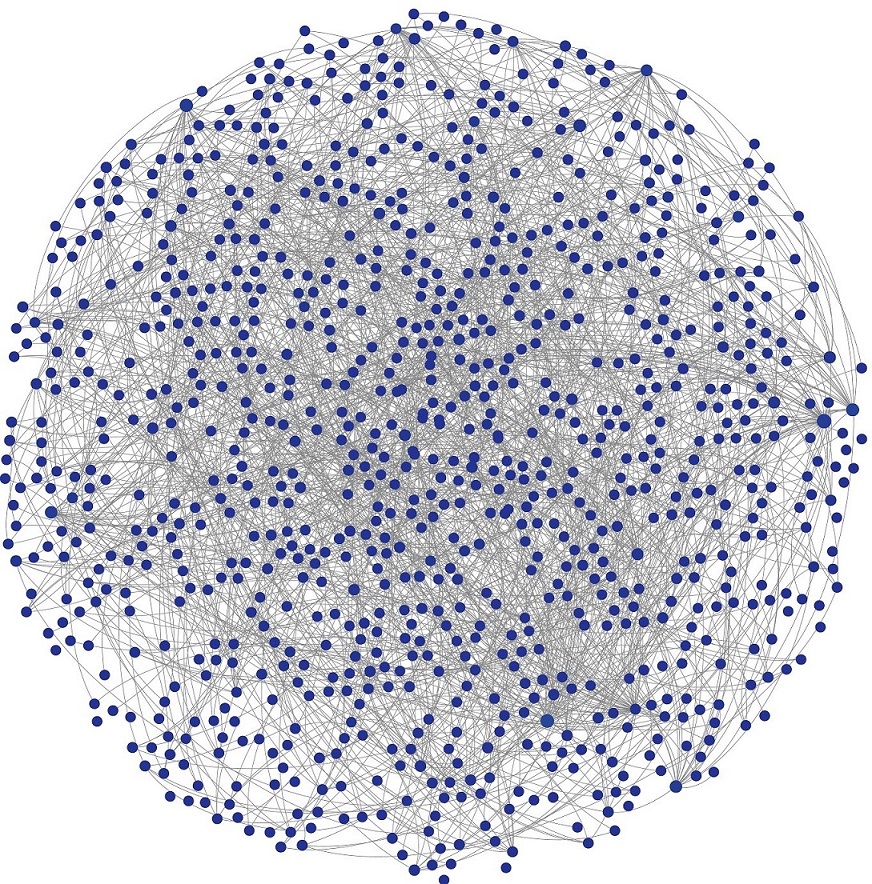}
\subcaption{1860s}\label{d1}
\end{minipage}%
\begin{minipage}{0.3\linewidth}
\includegraphics[width=0.82\linewidth]{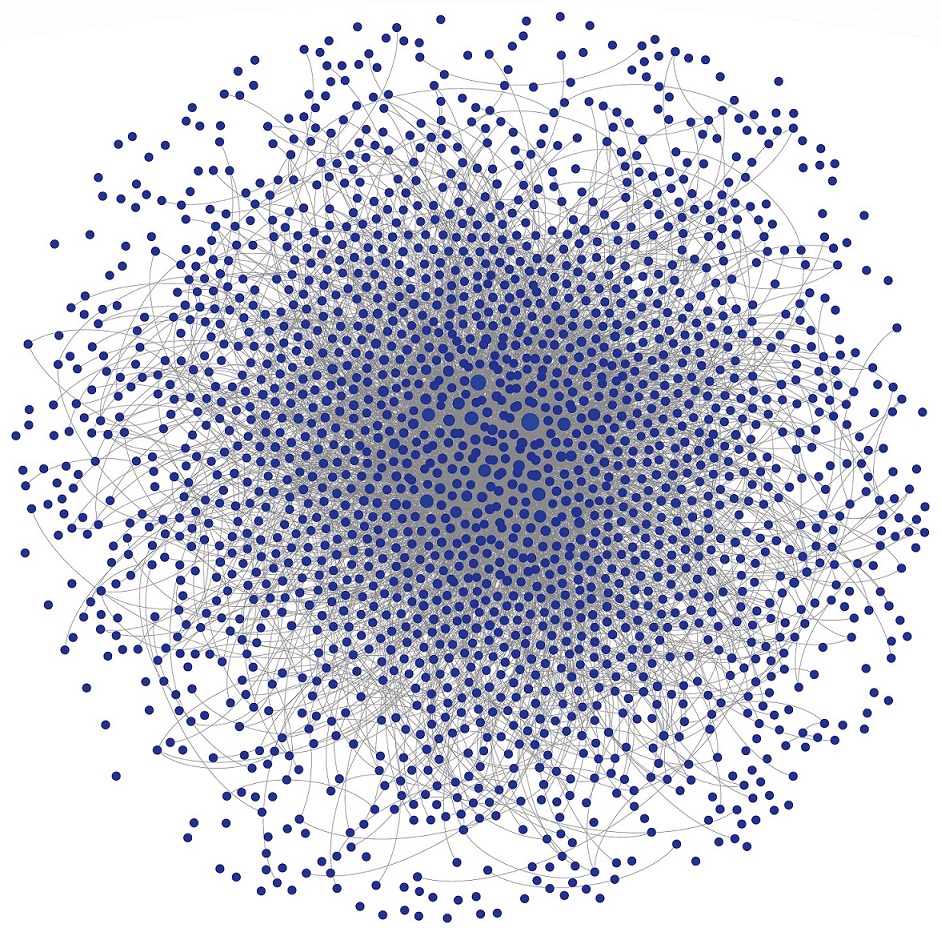}
\subcaption{1870s}
\end{minipage}%
\begin{minipage}{0.3\linewidth}
\includegraphics[width=0.82\linewidth]{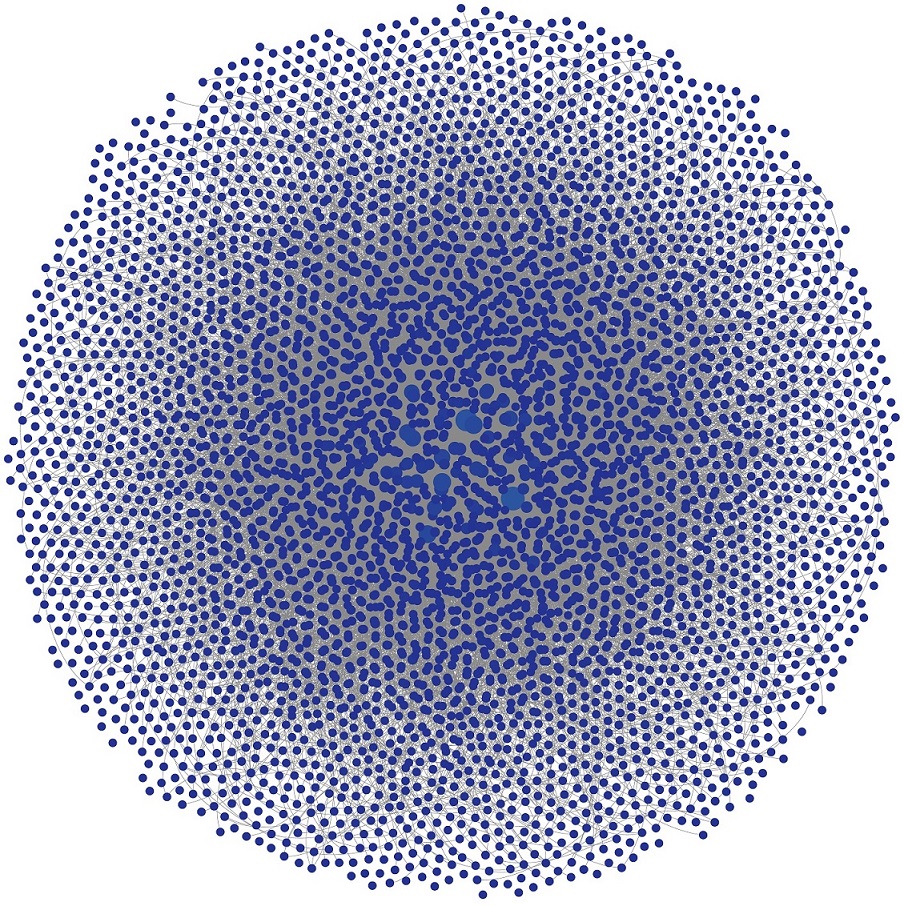}
\subcaption{1880-1909}
\end{minipage}%
\vspace{-\parskip}
\begin{minipage}{0.3\linewidth}
\includegraphics[width=\linewidth]{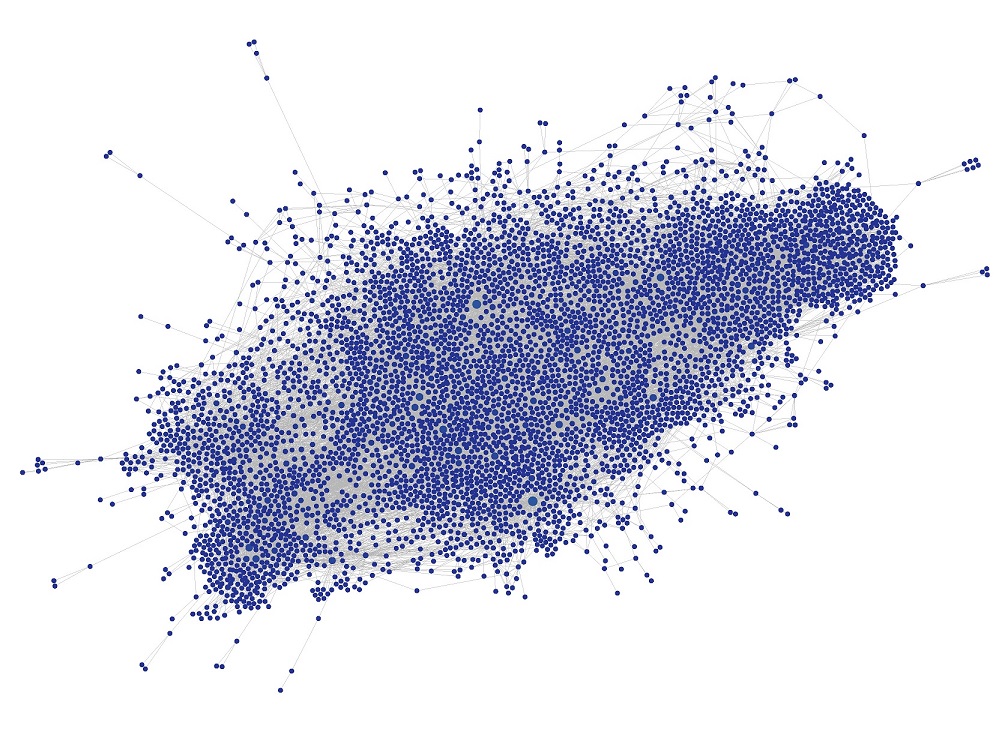}
\subcaption{1910s}
\end{minipage}%
\begin{minipage}{0.3\linewidth}
\includegraphics[width=\linewidth]{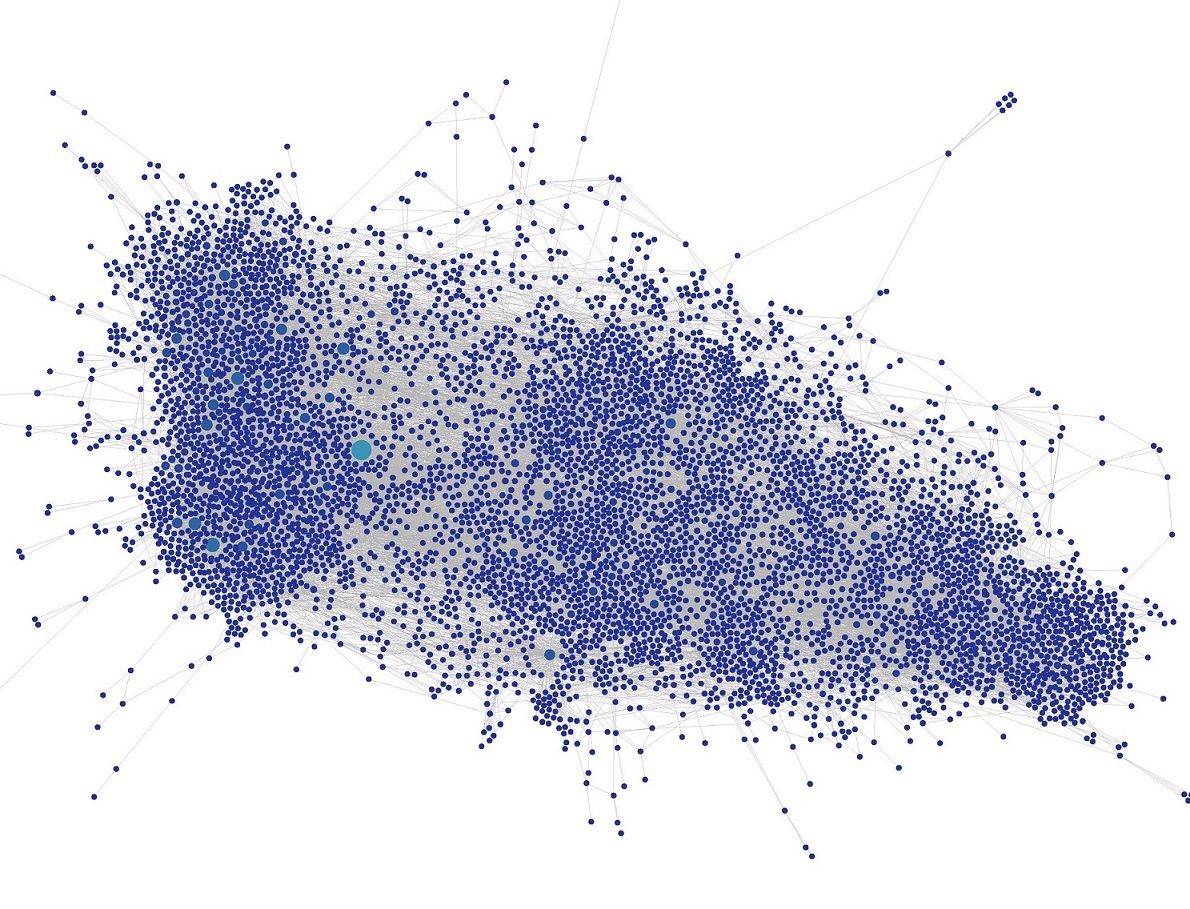}
\subcaption{1920s}
\end{minipage}%
\begin{minipage}{0.3\linewidth}
\includegraphics[width=0.9\linewidth]{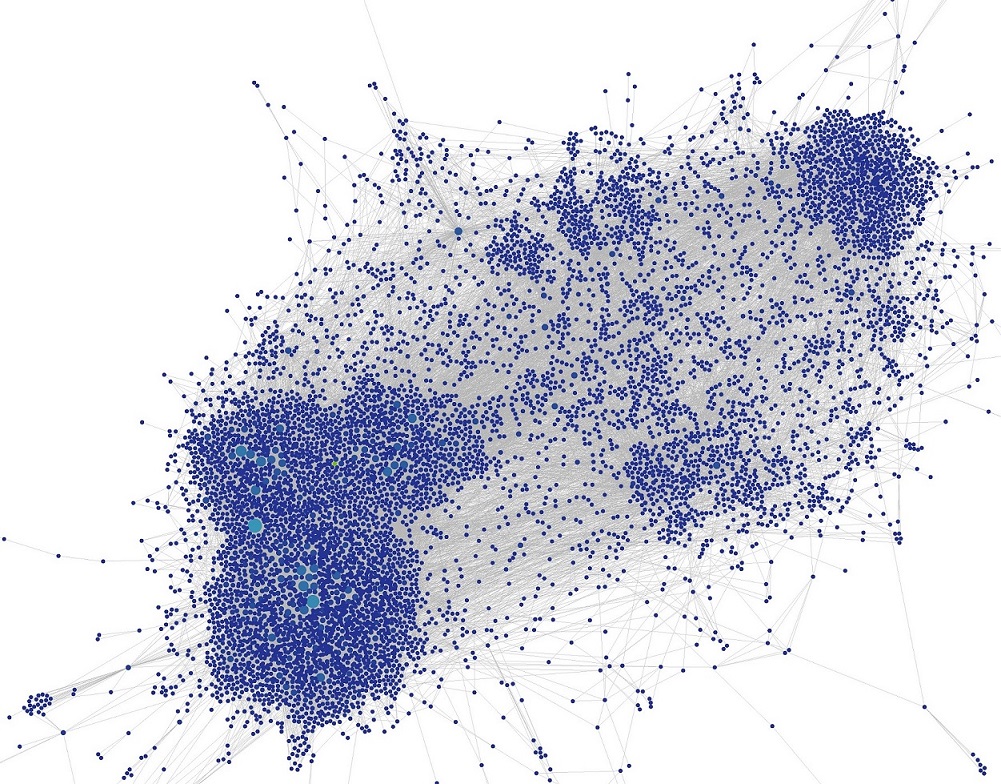}
\subcaption{1930-1959}
\end{minipage}%
\vspace{-\parskip}
\begin{minipage}{0.3\linewidth}
\includegraphics[width=\linewidth]{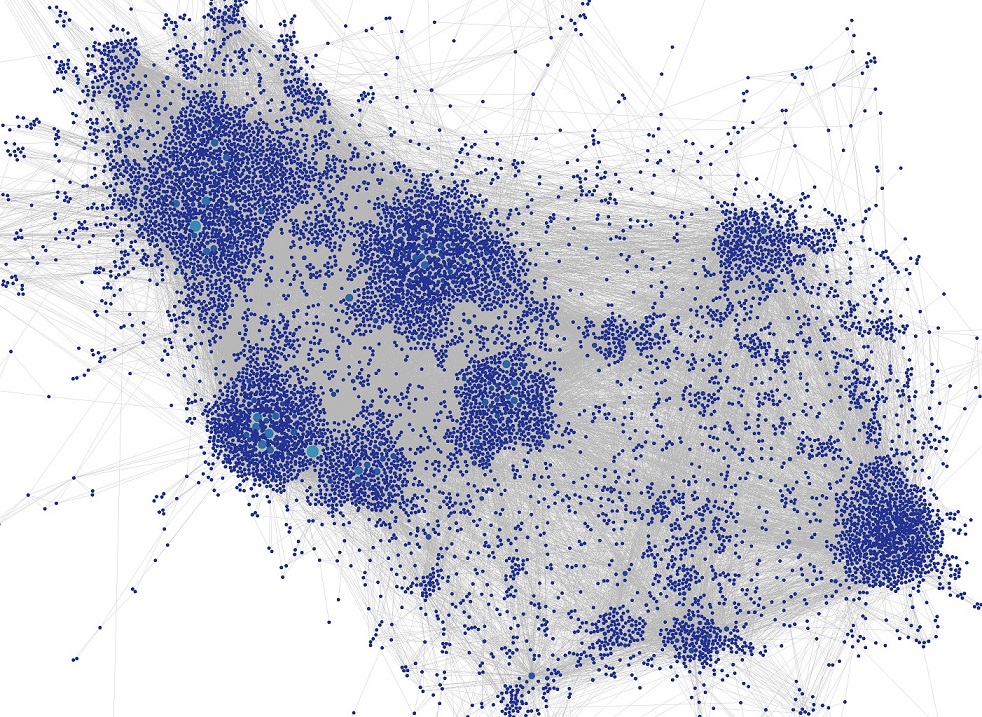}
\subcaption{1960-1979}
\end{minipage}%
\begin{minipage}{0.3\linewidth}
\includegraphics[width=.97\linewidth]{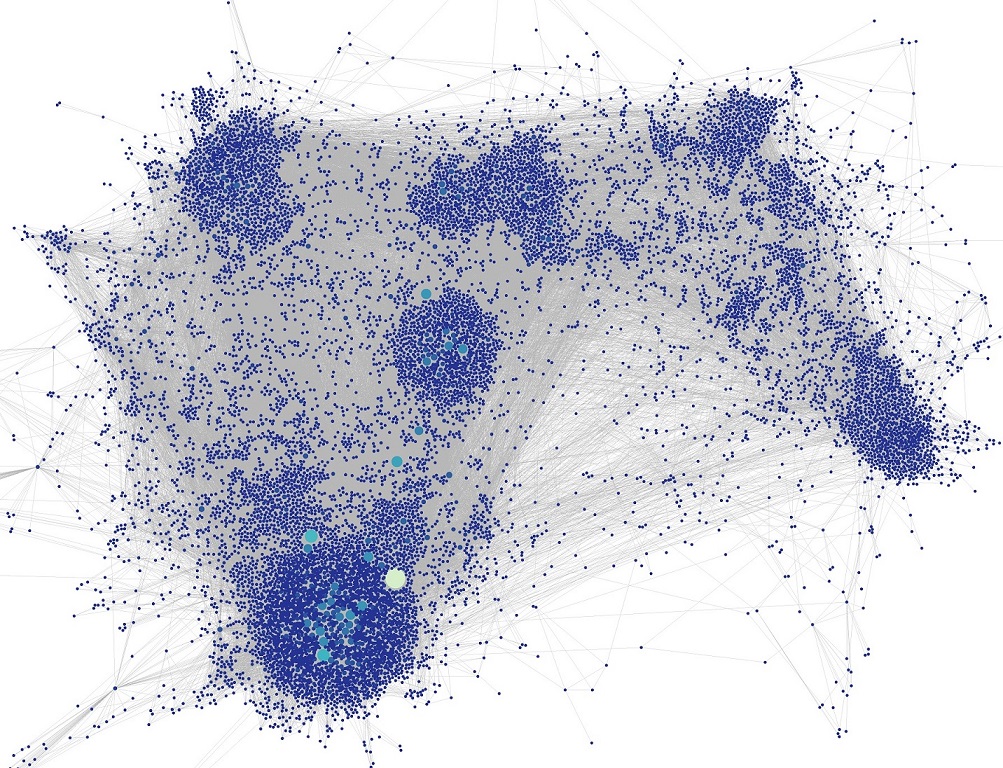}
\subcaption{1980-2009}
\end{minipage}%
\begin{minipage}{0.3\linewidth}
\includegraphics[width=\linewidth]{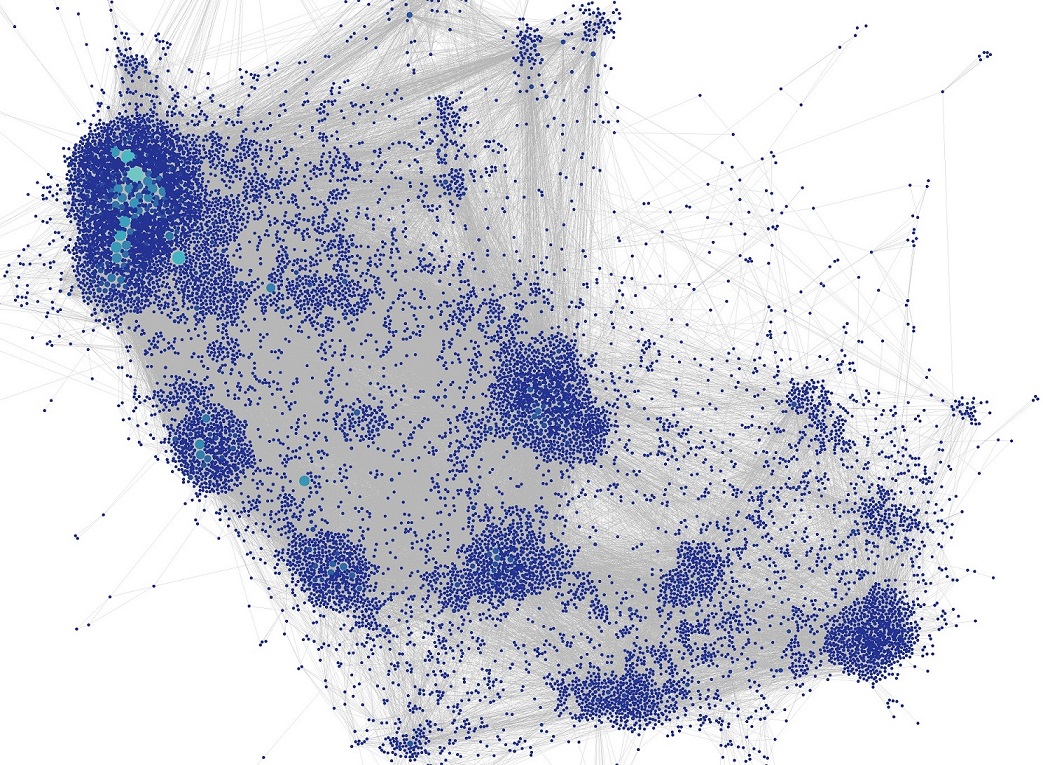}
\subcaption{2010s}
\end{minipage}%
\end{center}
\caption{Overview of the network structure evolution}\label{graph}
\end{figure}

\autoref{NewvsOld} compares the produced network based on the Information Extraction process with the earlier versions of the network that was built with parsing of limited available XML resources. As illustrated the network size and structure is significantly changed comparing to its earlier versions.
\begin{table}[h]
\centering{}\protect\caption{NZ Legislation Network, this study versus previous studies}
\label{NewvsOld}
\resizebox{\textwidth}{!}{\begin{tabular}{ l  l  l  l  l  l  l}
\hline
Network & Nodes & Edges & Average degree & Average CC\footnote{The average clustering coefficient (CC) is calculated based on the assumption that the network is directed using the approach that discussed in \cite{NSakhaee2016}} & Average path-length & Network type\\ 
 \hline
This study & 16385 & 137751 & 8.407 & 0.216 & 4.873 & dynamic\\ 
Previous studies & 3856 & 33884 & 8.878 & 0.39 & 3.569 & one snapshot\\ 
 \hline
\end{tabular}}
\end{table}
\autoref{graph} and \autoref{measures} capture the overall evolution of the Legislation Network in New Zealand from 1267 to the second quarter of 2018. To visualize the data, a network force-directed approach is used. 
In the layouts in \autoref{graph} each node is placed depending on their connection to the other nodes. As can be seen references between the Acts first appear in the 1840s, but the data-set visually looks like a graph since the 1850s and it gets denser from the 1870s. 
\begin{table}[h]
\centering{}\protect\caption{Overview of the network measures evolution}
\label{measures}
\resizebox{\textwidth}{!}{\begin{tabular}{ l  l  l  l  l  l  l  l  l  l  l  l  l}
\hline
Time & 1267-1839 & 1840s & 1850s & 1860s & 1870s & 1880-1909 & 1910s & 1920s & 1930-1959 & 1960-1979 & 1980-2009 & 2010s\\ 
 \hline
Number of nodes & 28 & 148 & 315 & 939 & 1945 & 4712 & 5473 & 6292 & 8622	& 11940 & 15524 & 16199\\ 
Number of edges & 0& 12 & 64 & 1252 & 4756 & 14851 & 18767 & 24538 & 44859 & 70683 & 121019 & 130969\\
Average degree & 0 & 0.081 & 0.203 & 1.333 & 2.445 & 3.152 & 3.429 & 3.900 & 5.203 & 5.920 & 7.796 & 8.085\\
Average path length & 0 & 1 & 1.046 & 2.605 & 3.592 & 8.061 & 7.514 & 8.301 & 6.164 & 5.554 & 5.051 & 4.927\\
Directed CC & 0 & 0 & 0.001 & 0.007 & 0.12 & 0.13 & 0.133 & 0.143 & 0.161 & 0.193 & 0.213 & 0.212\\
Small-world\footnote{The small-world sigma $\sigma$ is calculated by comparing clustering coefficient and average path length of each network to 50 equivalent random network with same average degree as suggested by \cite{Sigma}} $\sigma$  & NA & 0 & 0.447 & 1.165 & 15.587 & 75.173 & 82.519 & 80.122 & 24.239 & 16.131 & 195.837 & 208.084\\
 \hline
\end{tabular}}
\end{table}
As can be seen in \autoref{measures} the graphs show some small-world properties from 1860s with $\sigma>1$ and small-world property of the graphs is significant from 1970s comparing to 50 random graphs. As illustrated overlay the network gets denser and the average degree is growing. More significant clusters are observed during the most recent decades which can be seen in \autoref{graph}. These clusters could be the outcome of housekeeping activities such as edge and node removal, or it could be the result of mature referencing approach in legal drafting process. Both of the above hypotheses should be examined in future studies.
\begin{figure}
\begin{center}
\begin{minipage}{0.45\linewidth}
\includegraphics[width=\linewidth]{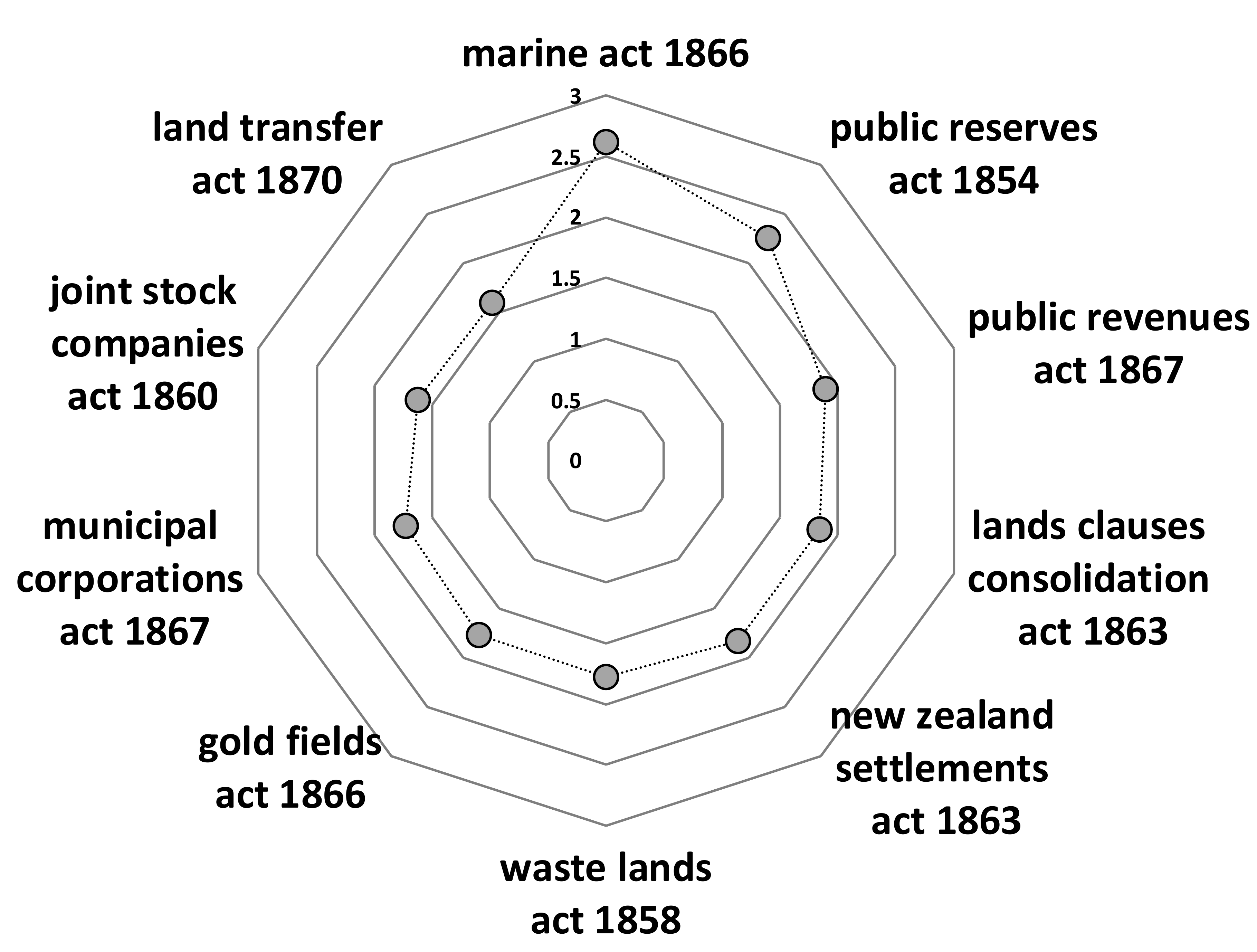}
\centering\includegraphics[width=0.6\linewidth]{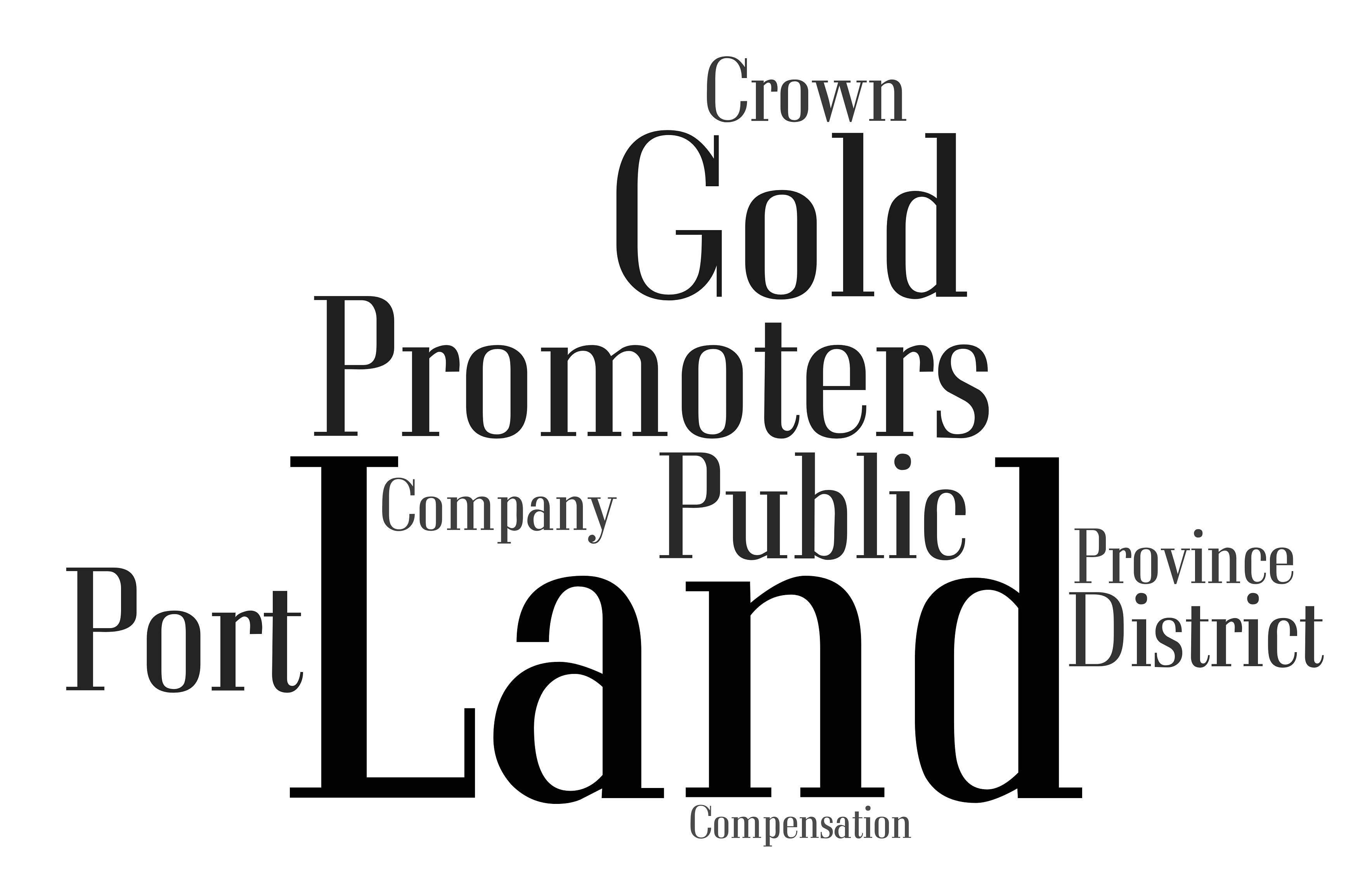}
\subcaption{1860s}\label{a2}
\end{minipage}%
\begin{minipage}{0.45\linewidth}
\includegraphics[width=\linewidth]{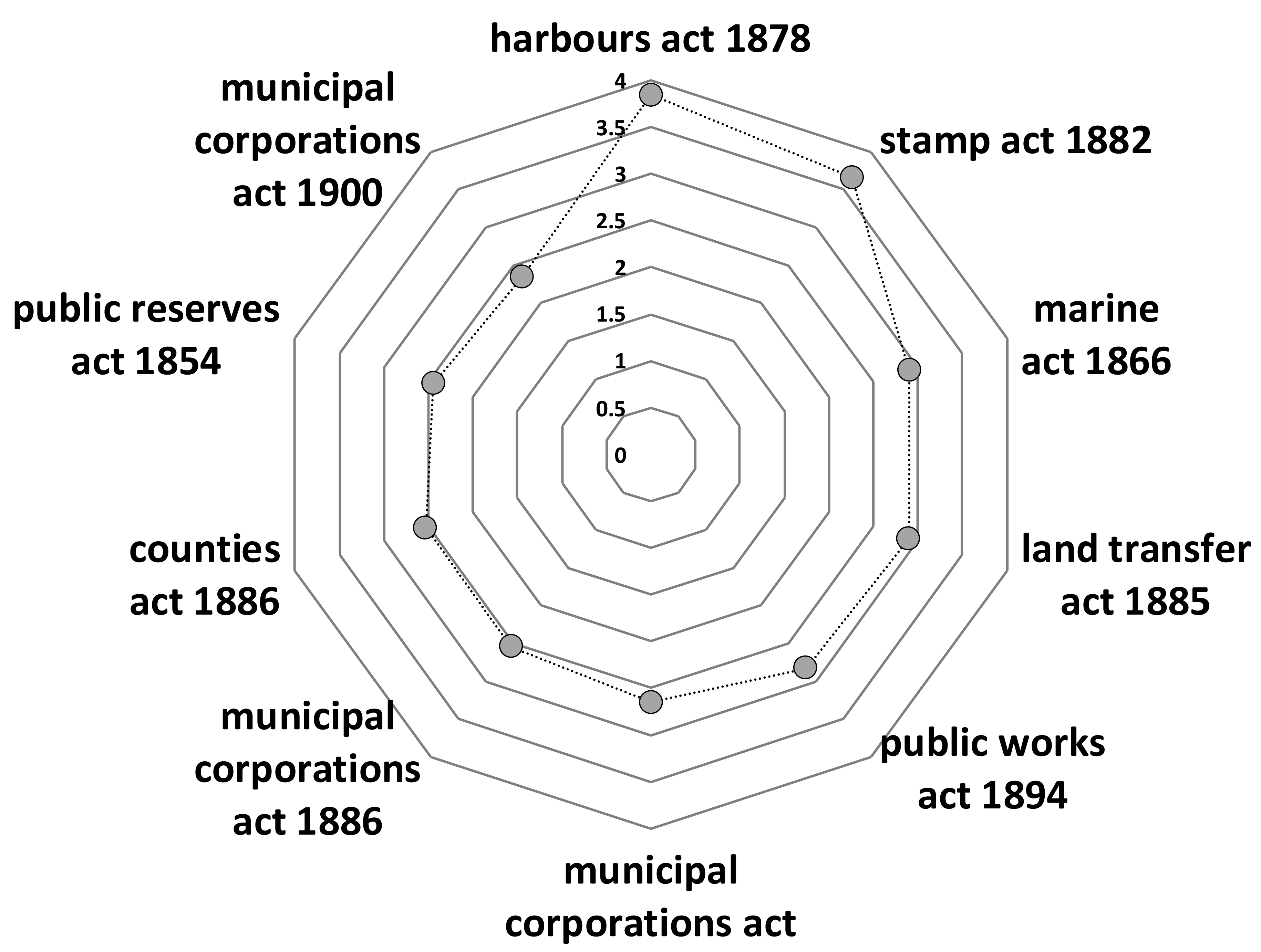}
\centering\includegraphics[width=0.6\linewidth]{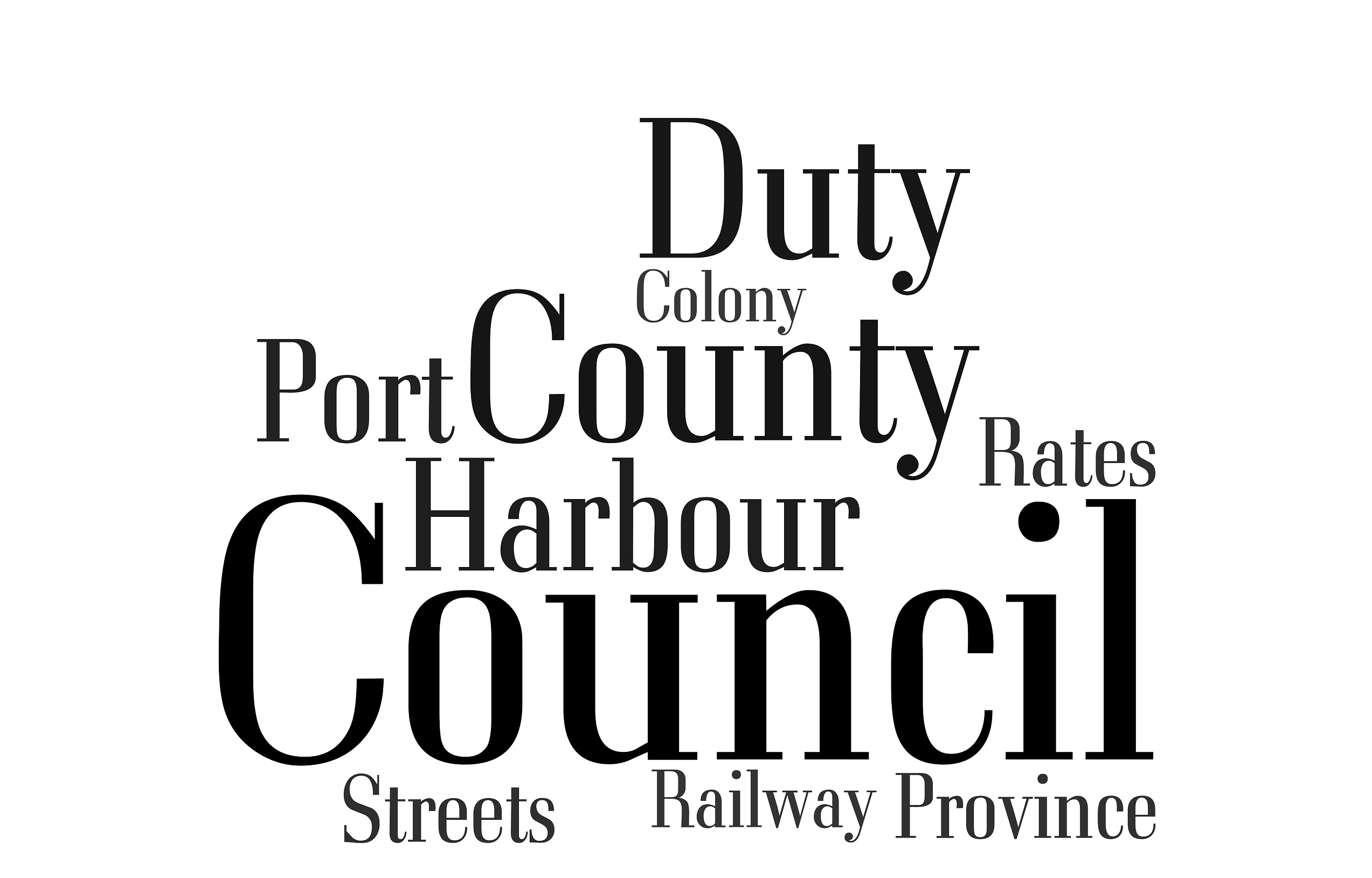}
\subcaption{1870-1909}\label{b2}
\end{minipage}%
\vspace{-\parskip}
\begin{minipage}{0.45\linewidth}
\includegraphics[width=.95\linewidth]{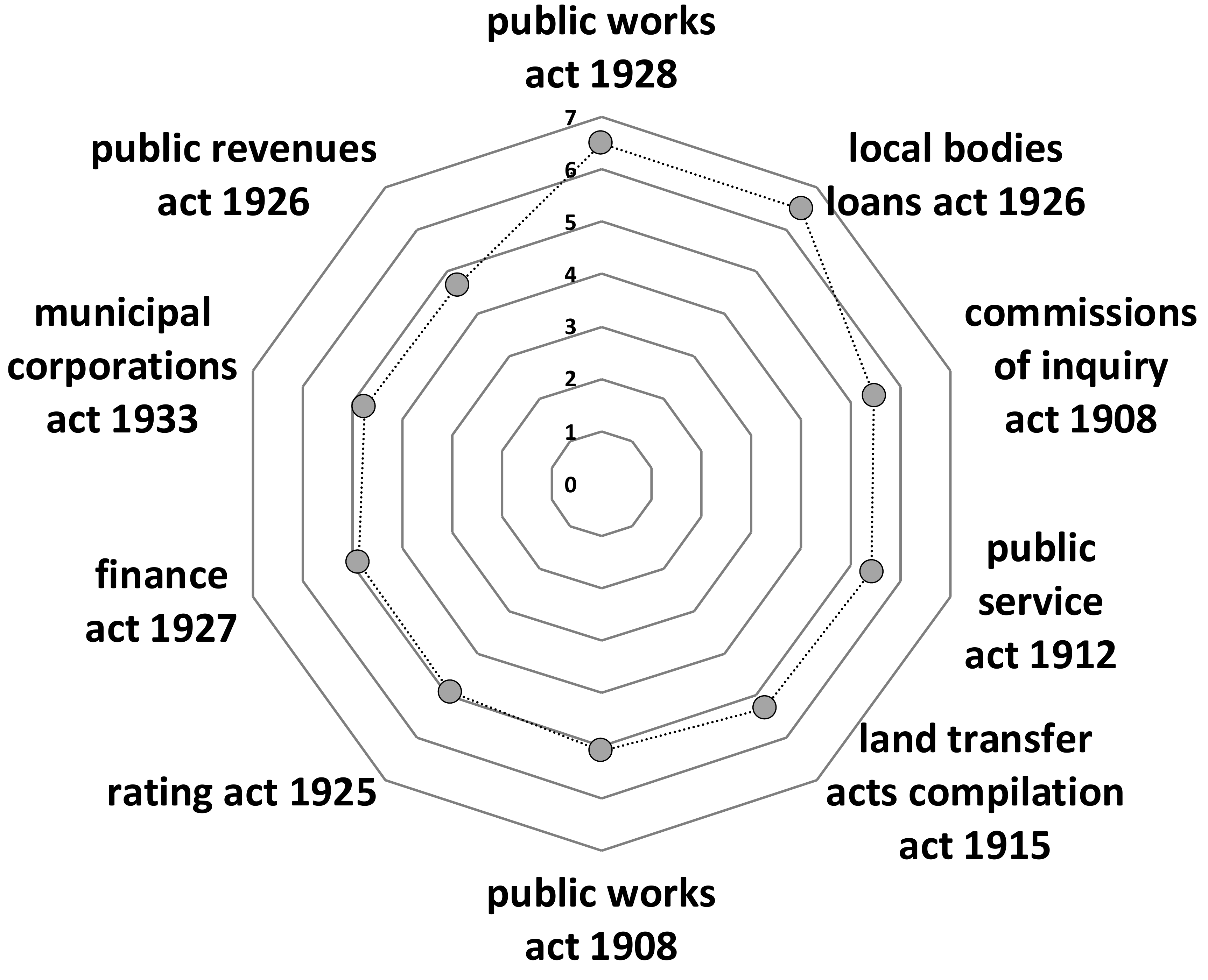}
\centering\includegraphics[width=0.6\linewidth]{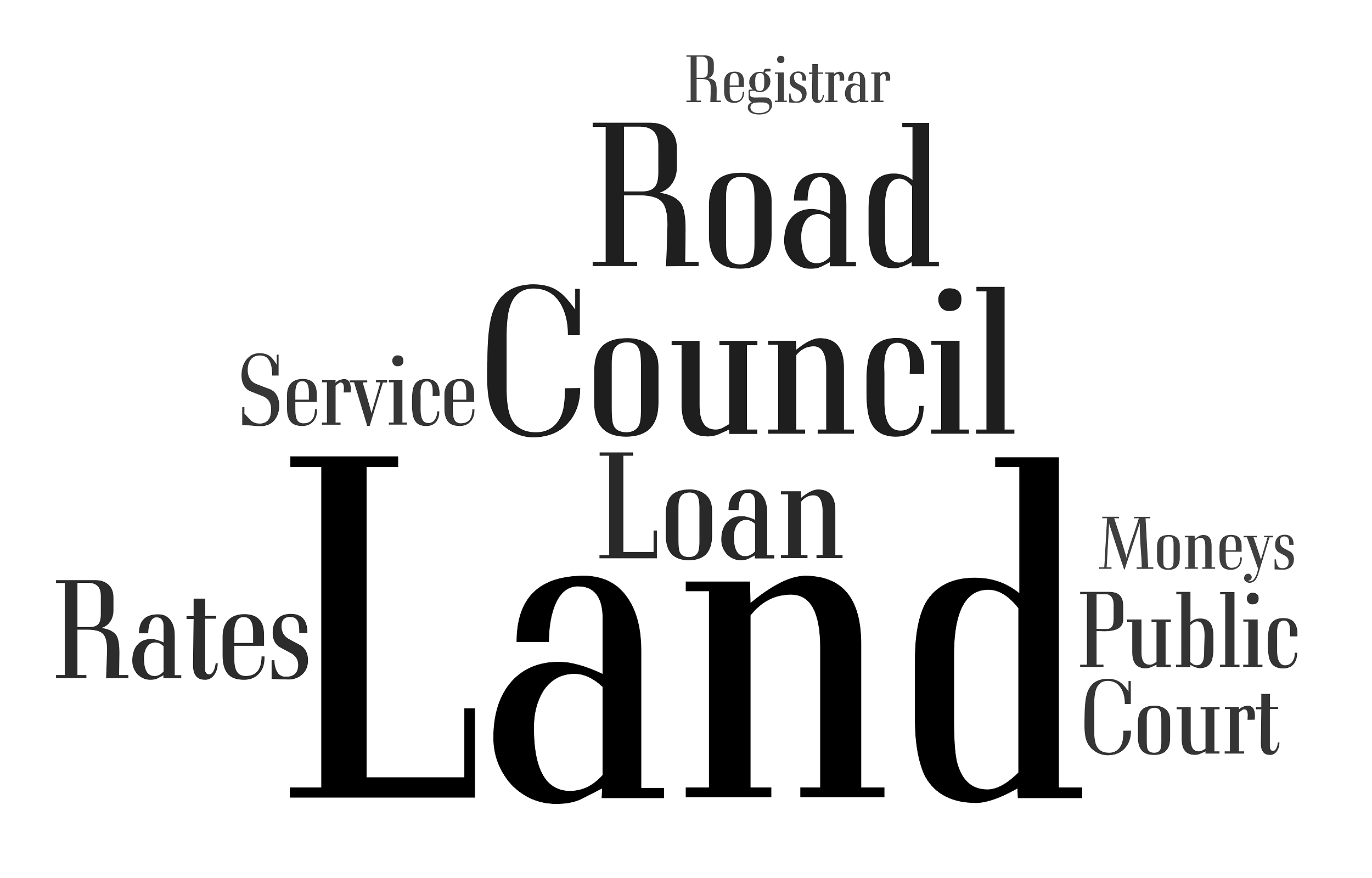}
\subcaption{1910-1959}\label{c2}
\end{minipage}%
\begin{minipage}{0.45\linewidth}
\includegraphics[width=\linewidth]{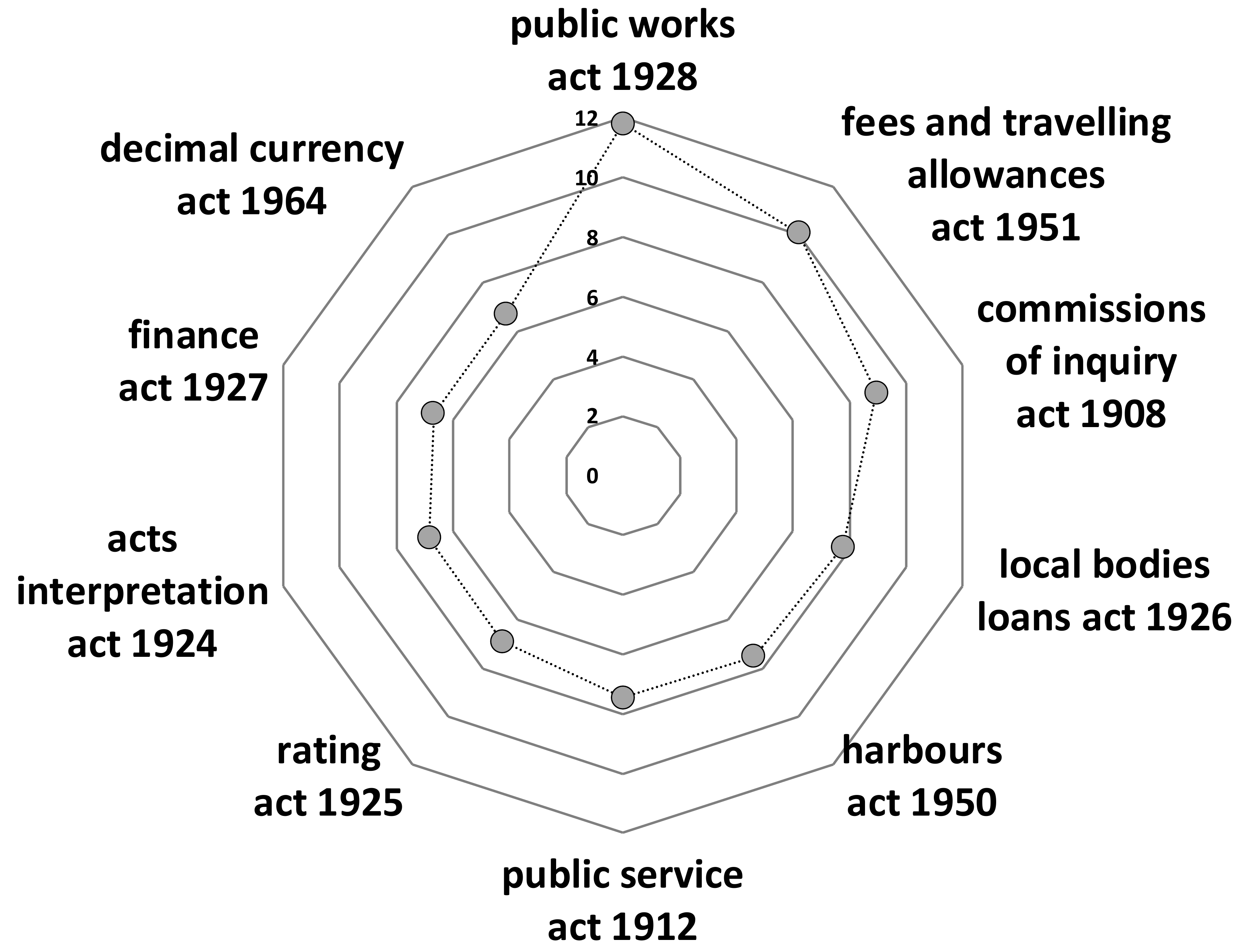}
\centering\includegraphics[width=0.6\linewidth]{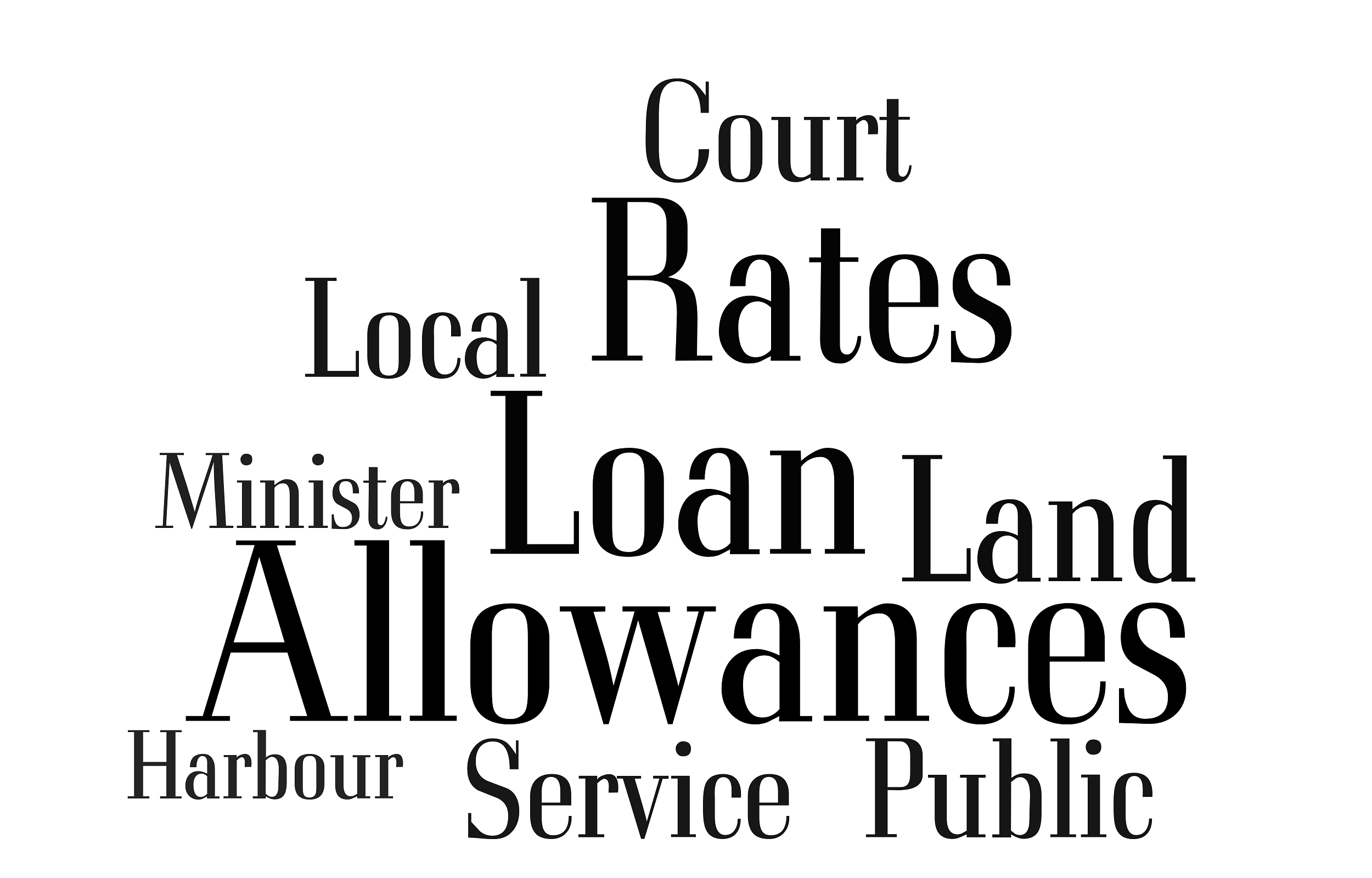}
\subcaption{1960-1979}\label{d2}
\end{minipage}%
\vspace{-\parskip}
\begin{minipage}{0.45\linewidth}
\includegraphics[width=\linewidth]{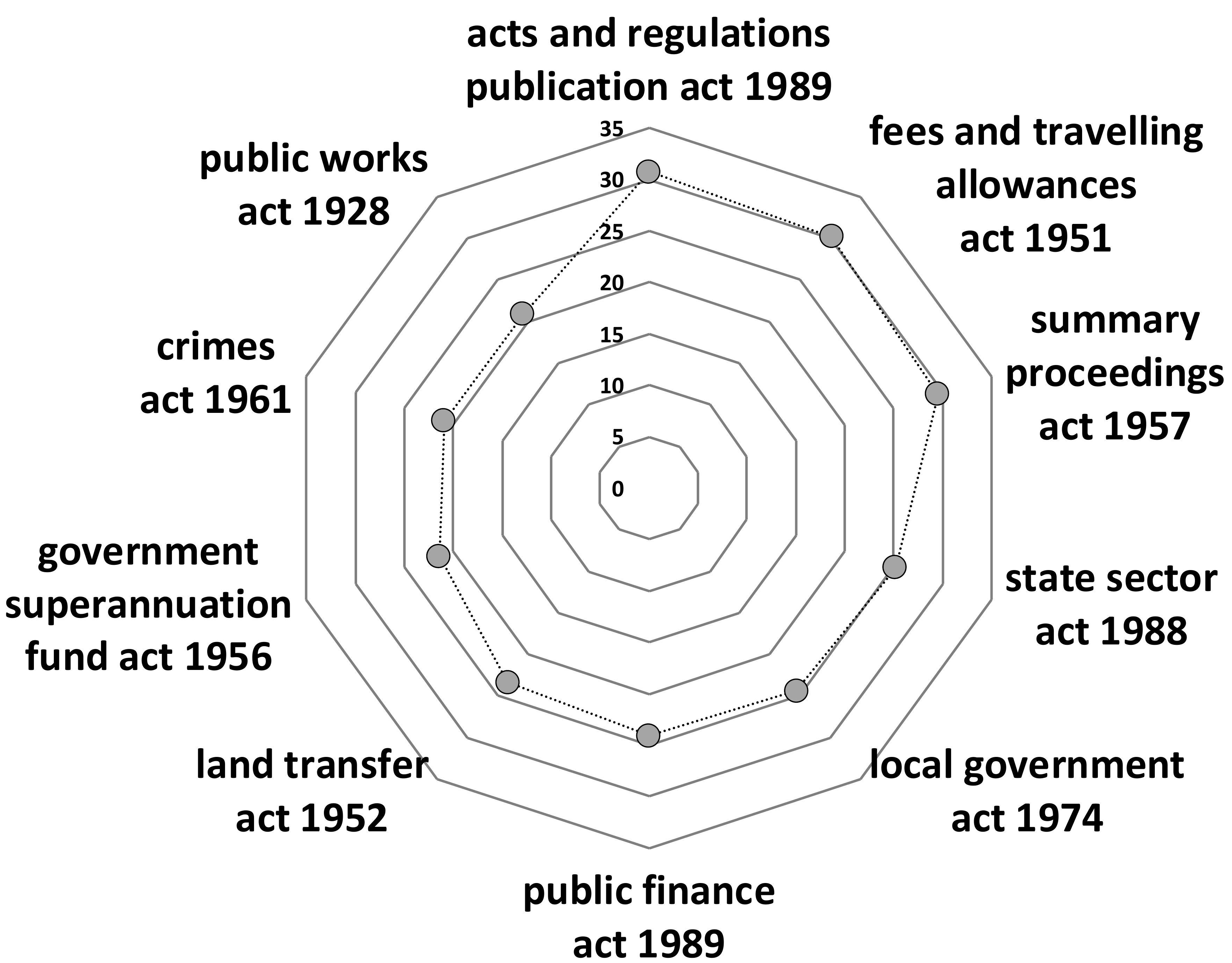}
\centering\includegraphics[width=0.6\linewidth]{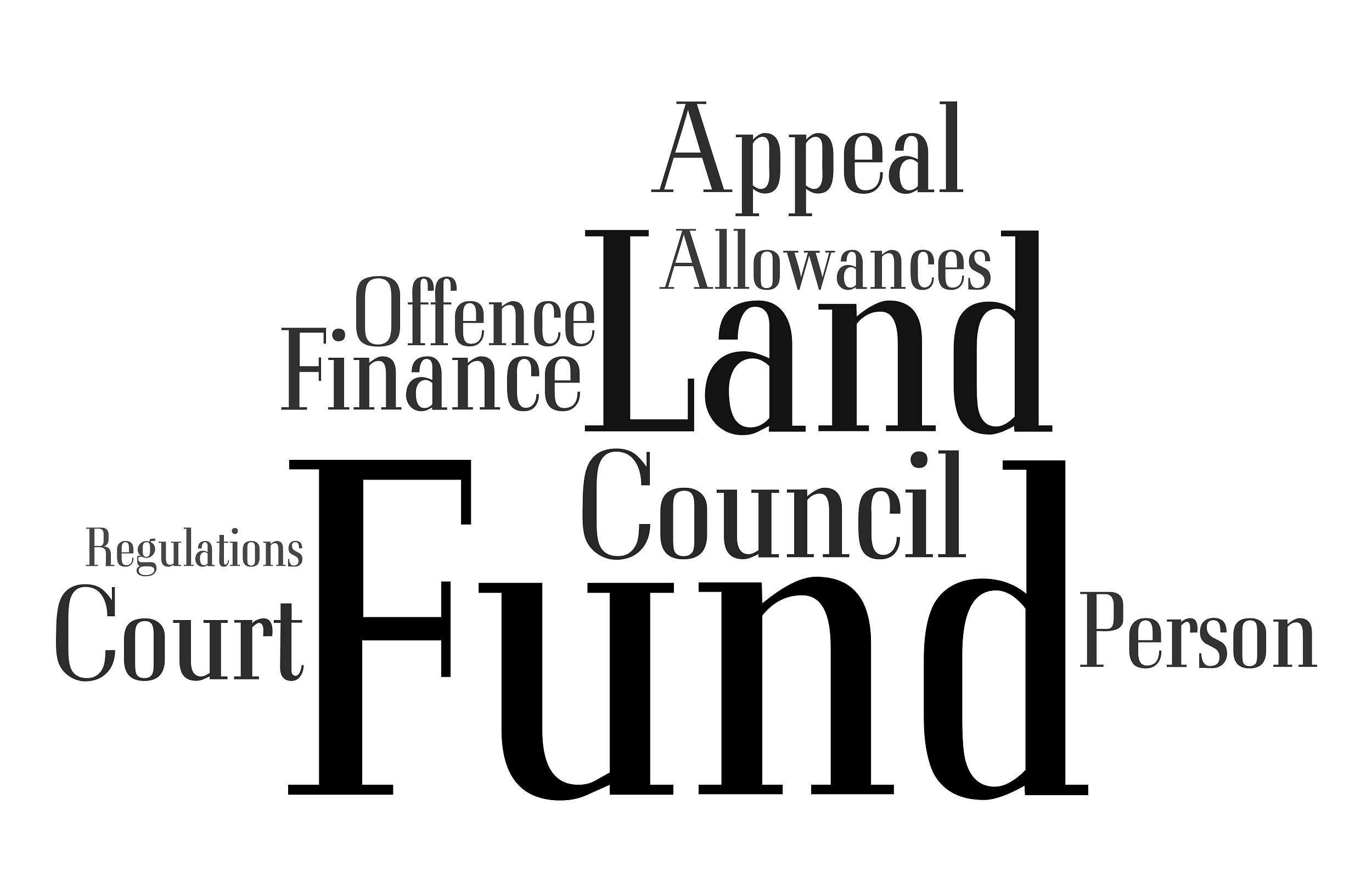}
\subcaption{1980-2009}\label{e2}
\end{minipage}%
\begin{minipage}{0.45\linewidth}
\includegraphics[width=\linewidth]{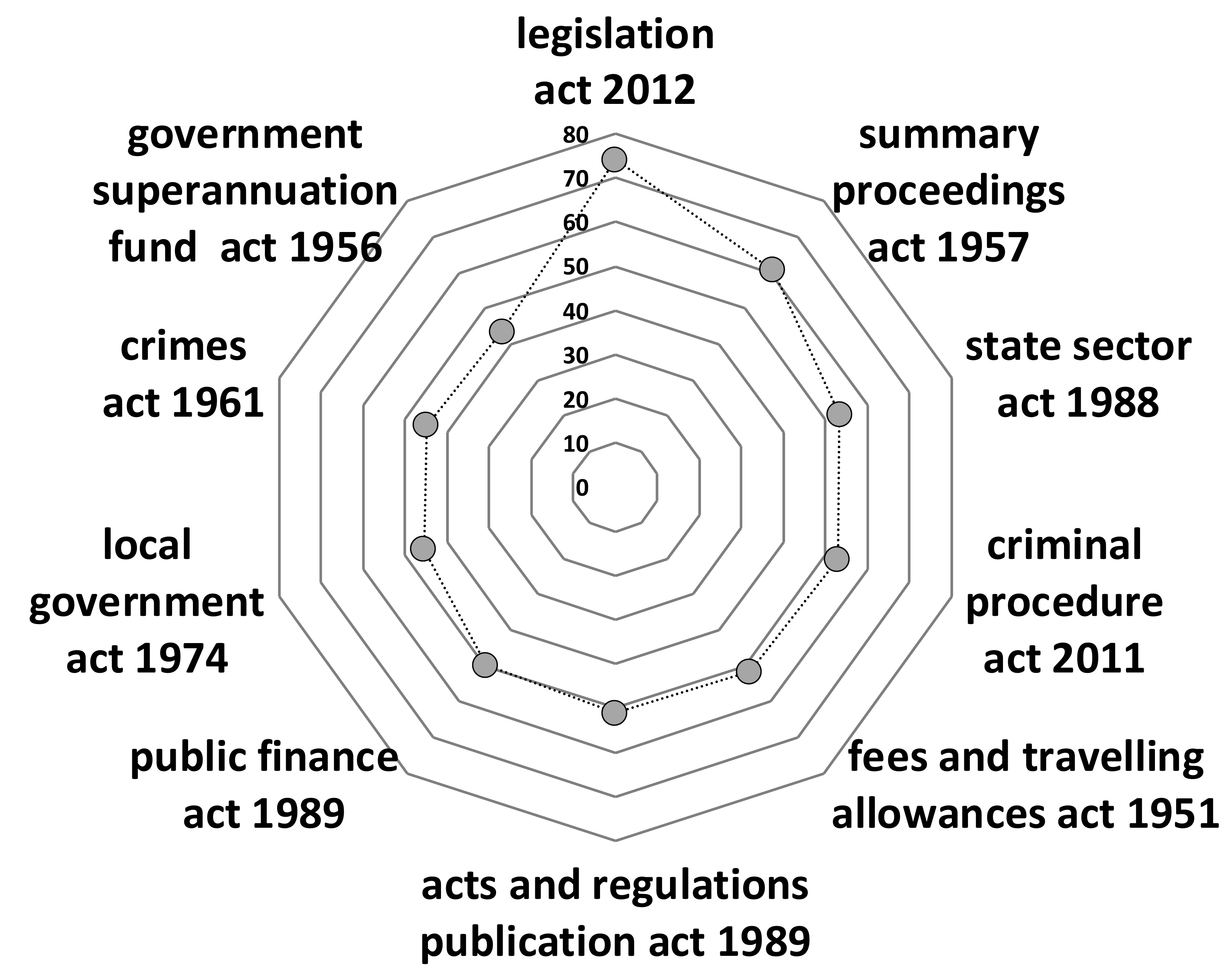}
\centering\includegraphics[width=0.6\linewidth]{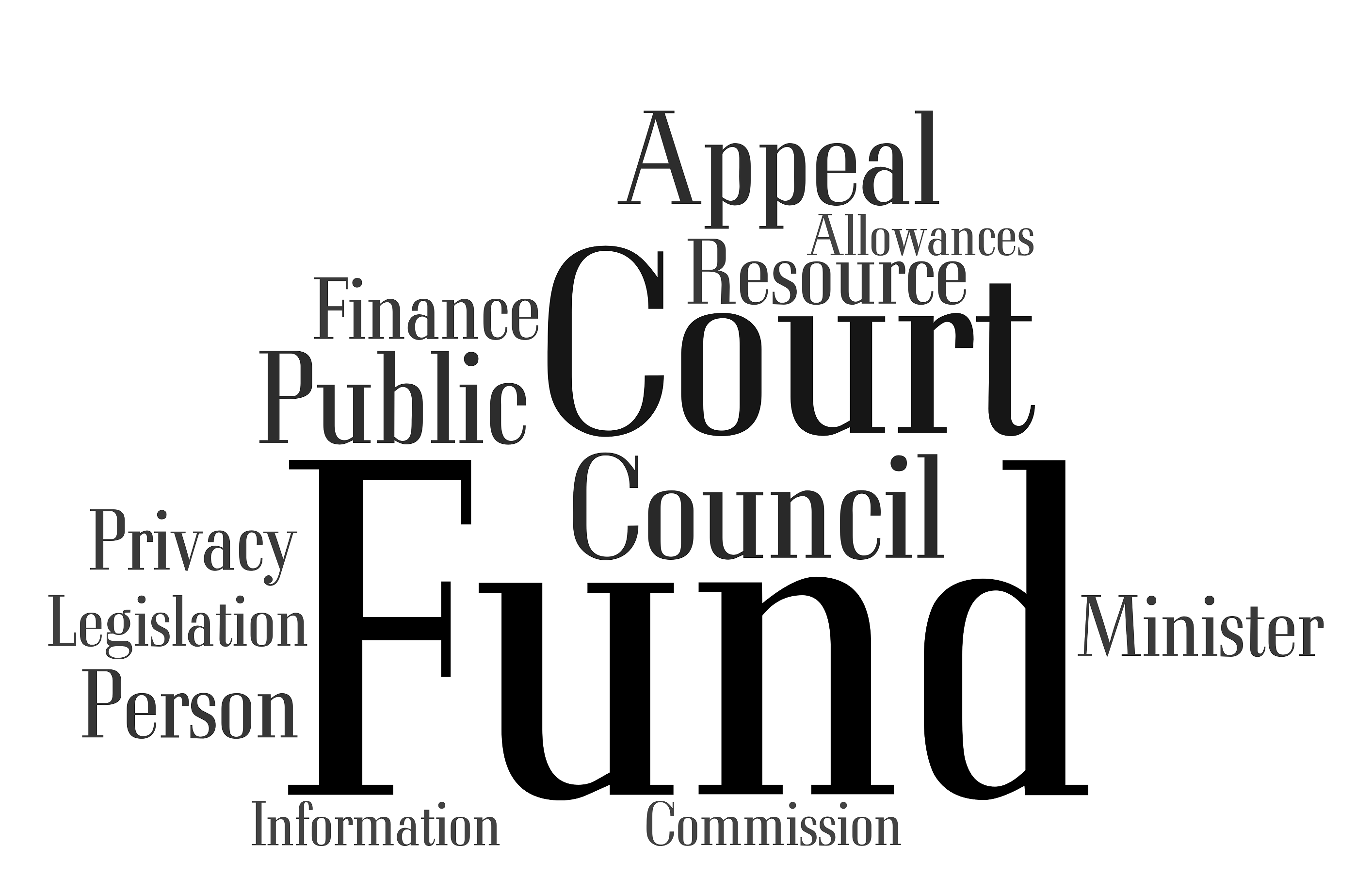}
\subcaption{2010-2018}\label{f2}
\end{minipage}%
\end{center}
\caption{Time evolution of the top ten legislation and the top subjects in the top twenty legislation}\label{chart}
\end{figure}

Based on the network structure information provided in \autoref{measures} and \autoref{graph}, six different time periods are chosen for the centrality evolution analysis. 
\autoref{chart} captures the time evolution of the top 10 nodes and the most frequent words \footnote{To find the frequent words, Textalyzer Python module is used. The frequent prepositions, conjunctions and articles are excluded from the analysis.} in the top 20 nodes based on Katz prestige centrality measure. 

As mentioned earlier, prior to the 1860s the graphs don't show significant small-world properties. The visual presentation in \autoref{a1} to \autoref{c1} also reflects that the network can be considered as a random graph during this period. So the Katz centrality degree distribution is nearly a uniform distribution in these time periods and is excluded from \autoref{chart}.

\autoref{a2} shows the most important nodes with the impression that \textbf{Land} was the most important law subject back at that time.  In the next selected time period the network shows small-world properties, and as can be seen in \autoref{b2} the centrality measure shows a higher kurtosis with the word \textbf{Council} being the most frequent topic in legal domain. 

Similarly the other graphs reflect the change in the network structure and highlights the relationship between the laws and the socio-economic requirements of the country. In the current decade, with the new sets of legislation being introduced and referenced to the older documents, the centrality measure is increased comparing to the previous decade, and the hot legal topics show a change which could be a good reflection of the society's needs.
\section{Evaluation and Robustness}
In this section the performance of the proposed framework is discussed. As explained in the previous section, the main goal of the study is to extract the information to build Legislation Network. The framework includes Named Entity Recognition, Relation Extraction, and Approximate String Matching jointly to extract the network's node and edge information. In this section the proposed framework is evaluated and the related errors are calculated. The familiar metrics of recall and precision measures are used to evaluate the system. High \textbf{precision} means that the framework returns substantially more relevant results than irrelevant ones, while high \textbf{recall} means that the process returns most of the relevant results. At the end of this section the impact of the identified errors on the network structure is explained, and the robustness is assessed.
\subsection{Error Estimation, Precision, and Recall}
\begin{figure}[h]
\centering
\includegraphics[scale=1]{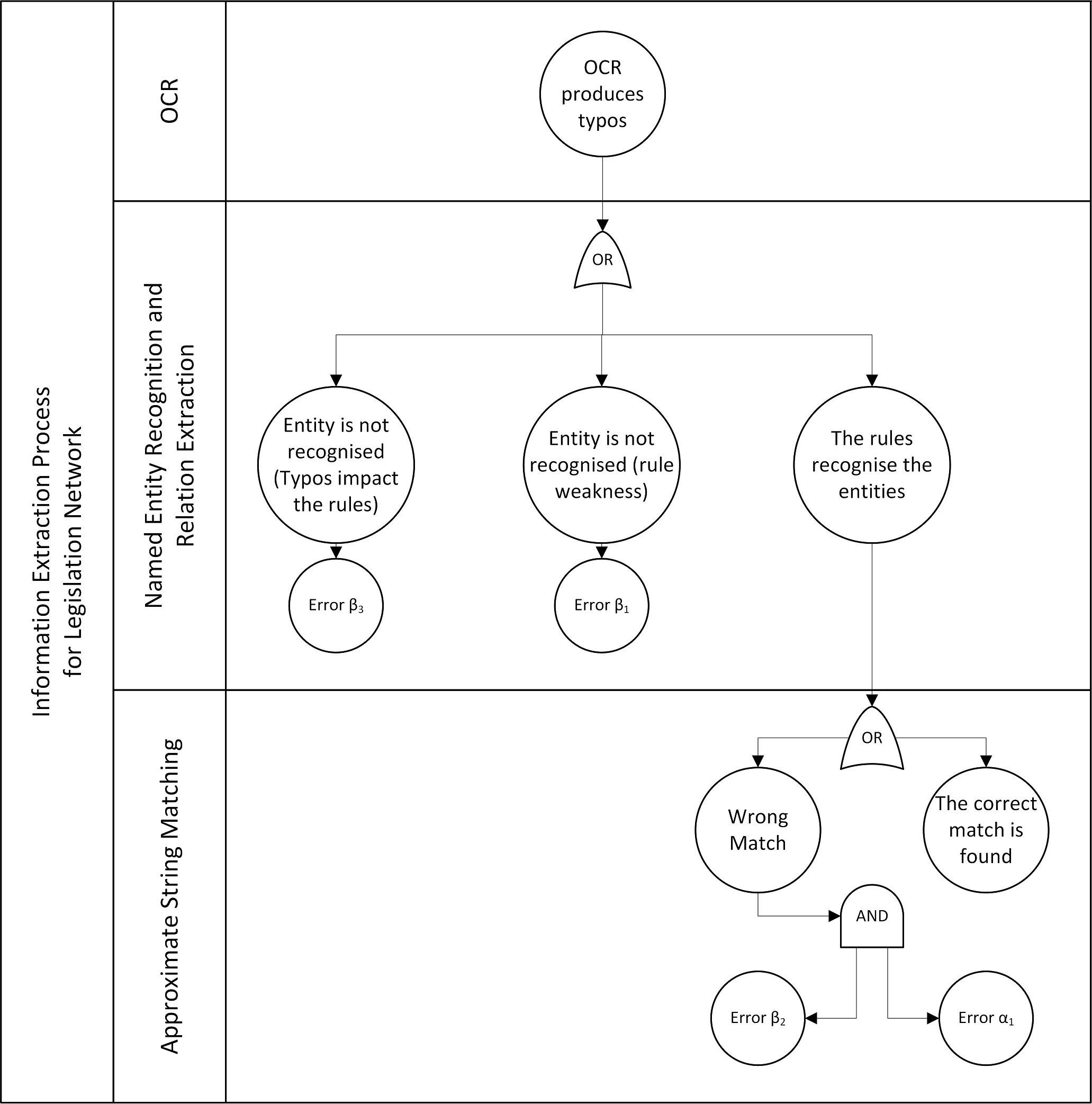}
\caption{Error Diagram}
\label{Error}      
\end{figure}

In the proposed Information Extraction process, Named Entity Recognition is combined with Approximate String Matching to recognize, validate and optimize the entities (the nodes and the edges).
\autoref{Error} illustrates the occurrence of the false-positive and false-negative errors in this process and helps in studying the robustness of the network. 

If the Entity Recognition process finds the entities, then there is a possibility that the Approximate String Matching process fails to find the correct match. A type \textit{I} error $\alpha_1$ occurs when the approximate string matching process fails to find the correct match. From one side this issue contributes to the false-positive error because it adds invalid entities to the output. These invalid entities impact the accuracy of the node list and the edge list of Legislation Network. To estimate the $\bar\alpha_1$ in the case study, a cluster sampling method is used to randomly choose ten sets of 30 entities. By manual check of the samples, the rate of incorrectly matched entities is observed. A Kolmogorov-Smirnov test suggests that the estimated error $\bar\alpha_1$ has a normal distribution with the parameters in \autoref{errorrates}.

$\beta_1$ also occurs when approximate string matching system picks a wrong match for the entities. This issue can contribute to the false-negative error because those entities that are wrongly matched to other entities are missing from the data set. The estimation methodology and the estimated value of $\bar\beta_1$ are equal to that of $\bar\alpha_1$ as indicated in \autoref{errorrates}.

$\beta_2$ is measuring the Information Extraction rules' performance. If it fails to recognize entities, then those entities are missed, and it results in another type of false-negative error. The estimation process for $\bar\beta_2$ is different from the previous two errors, and it is harder to address. For the case study a sample set of 30 text files are randomly chosen using cluster sampling method. Then all of the extracted entities for each document is compared to the actual entities in a human involving process. The list of missing entities is categorized into two parts: caused by a typo, or caused by insufficient rules to recognize the entity. The rate of missing entities caused by weak or missing rules is calculated for each document and denoted by $\bar\beta_2$. The Kolmogorov-Smirnov results through all of the 30 documents show that $\bar\beta_2$ has a normal distribution with the parameters in \autoref{errorrates}.\\~

\begin{table}[h]
\centering{}\protect\caption{Errors, sensitivity and specificity}
\label{errorrates}
\renewcommand{\arraystretch}{2}
\begin{tabular}{ l  l  l  l  l | l  l  }
\hline
Measure & $\bar\alpha_1$ & $\bar\beta_1$ & $\bar\beta_2$ & $\bar\beta_3$ & $\bar\alpha$ & $\bar\beta$\\ 
\hline
 $\mu$ & 0.0160 & 0.0160 & 0.0012 & 0.0007 & 0.0160 & 0.0179\\ 
 $\sigma$ & 0.0012  & 0.0012 & 0.0001 & 0.0001 &0.0012 & 0.0012\\ 
 \hline
\end{tabular}
\renewcommand{\arraystretch}{1}
\end{table}
$\beta_3$ addresses the error when typos cause problem for recognizing the entities. The estimation process is very similar to that of $\bar\beta_2$. A sample of 30 text files are collected. Then the rate of missing entities caused by OCR typos is calculated for each document and addressed as $\bar\beta_3$. The Kolmogorov-Smirnov results through the selected documents show that $\bar\beta_3$ has a normal distribution with the parameters in \autoref{errorrates}. In the sample it is observed that the typos that cause the entity recognition failure are only numeric typos. For example OCR might produce an error and convert 1987 to l987 by misspelling number 1 to letter l. Then the Information Extraction rules are impacted to recognize l987 as a year, so the entity is missed.\\~

As \autoref{Error} shows $\alpha_1$ is the only false positive error which contributes to the \textbf{overall false positive error} of the system. \autoref{errorrates} captures $\bar\alpha$, assuming that $\bar\alpha$ estimates the overall type one error.
To estimate the \textbf{overall false negative error} of the system, $\beta_1$, $\beta_2$, and $\beta_3$ are considered as mutually exclusive events. From \autoref{Error} it is also clear that the intersections of each two of these errors are empty, so they are independent. \autoref{errorrates} shows $\bar\beta$, the estimated value for the overall false negative, or the type \textit{II} error.\\~

To calculate the Precision and Recall, \autoref{precision} and \autoref{recall} are used. 
\small
\begin{flalign}\label{precision}
    Precision &= \frac{True Positive}{True Positive+False Positive}= \frac{1-\bar\alpha-\bar\beta}{1-\bar\beta}
\end{flalign}
\begin{flalign}\label{recall}
    Recall &= \frac{True Positive}{True Positive+False Negative}= \frac{1-\bar\alpha-\bar\beta}{1-\bar\alpha}
\end{flalign}
\normalsize

Referring to the above equations and  \autoref{errorrates}, the Precision of 98.37\% and Recall of 98.18\% are calculated. These outcomes suggest the high performance of the Proposed Information Extraction framework that results in a high data reliability of the output Legislation Network. 

As explained in section \ref{Approximate String Matching}, the proposed hybrid Approximate String Matching technique substantially reduces the errors. It is important to mention that at the earlier stages of the study by using the classic string matching techniques, the error rates were considerably higher, and the accuracy of the network was questionable. A time consuming examination process engaging manual checks was applied to propose the hybrid model which resulted in impressive performance and high precision and recall. The improvement obviously involved a lot of efforts and time, but resulted in accuracy and confidence in Legislation Network studies.

\subsection{Robustness}
With a coherent understanding of the errors, it is very important to study the robustness of the network to the error. The robustness study proves the importance of the data accuracy which supports the value of the proposed hybrid model for the approximate string matching. In this section to study the network robustness, diameter and three major centrality measures are used.

\begin{figure}[h]
\centering
\includegraphics[scale=.4]{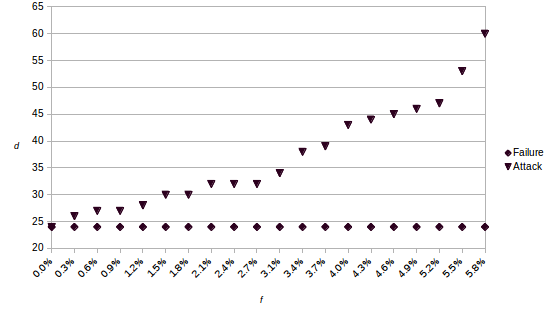}
\caption{Changes of the network diameter $d$ as the function of fraction $f$}\label{Robust1}
\end{figure}

To understand the diameter robustness of the network, attack and failure analysis is required. As discussed earlier Legislation Network in general show scale-free characteristics. So it is expected to observe a reasonable error tolerance of the network as the result of random failures, but vulnerability as the result of attacks\cite{robust2}. To study the network robustness to node failures the method is to randomly remove a fraction of nodes $f$ and recalculate the diameter of the network $d$. To study the network robustness to attack by removing a fraction $f$ of the largest nodes \footnote{Based on their connectivity (total degree)} and observe the change to the the diameter $d$. The results of both failure and attack to the nodes are captured in \autoref{Robust1}.
The observed tolerance to failures and the vulnerability to attacks shows that the connectivity is provided by a few highly connected nodes, and majority of nodes have only few edges. 
As can be seen the vulnerability to the attacks starts immediately after removing a small fraction $f = 0.3\%$ of the highly connected nodes. This scenario of attack is highly unlikely in Legislation Network considering the high Precision and Recall of the proposed data extraction process.

As discussed in the previous studies \cite{NSakhaee2016} \cite{NSakhaee2017}, the most relevant centrality measure for Legislation Network is the Katz second prestige measure. In recent studies,  the reliability of different centrality measurements against network manipulation has been addressed \cite{CRobust2} \cite{CRobust}, but Katz prestige centrality is not much discussed. In this paper the Katz centrality, betweenness centrality, and degree centrality robustness of Legislation Network against edge deletion error is studied. 
To address the robustness, four major measures of accuracy that proposed in \cite{CRobust2} and \cite{CRobust} are used. These measures are Top 1, Top 3, Top 10 percent, and the Pearson correlation to compare the centrality measures between the true network and the manipulated network.

\begin{figure}[h]
\begin{minipage}{0.35\linewidth}
\includegraphics[width=\linewidth]{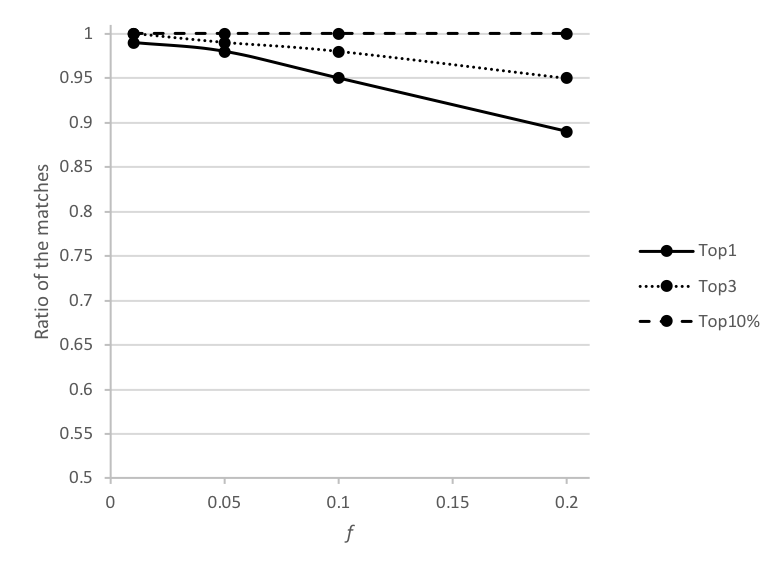}
\subcaption{Katz Prestige Centrality}
\end{minipage}%
\begin{minipage}{0.35\linewidth}
\includegraphics[width=\linewidth]{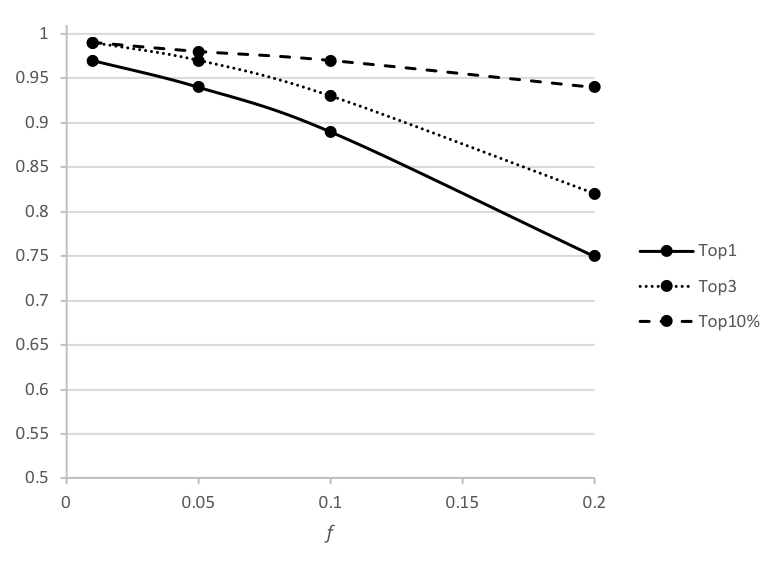}
\subcaption{Betweenness Centrality}
\end{minipage}%
\begin{minipage}{0.35\linewidth}
\includegraphics[width=\linewidth]{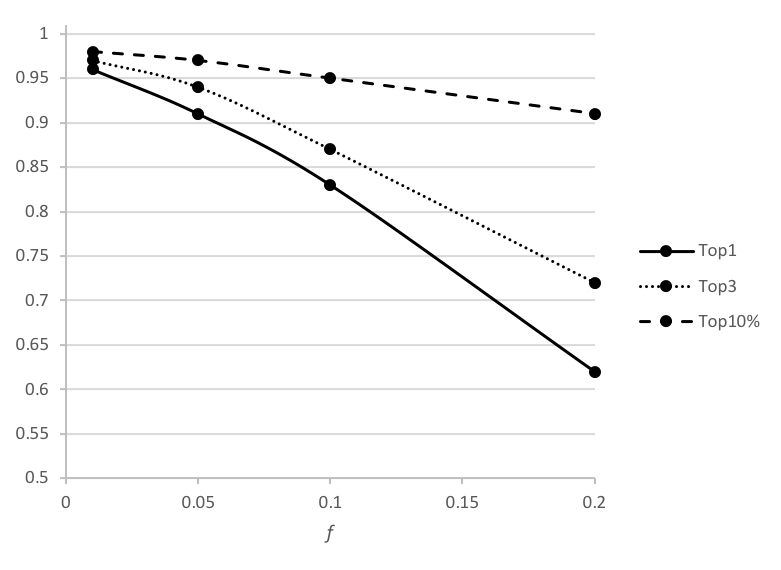}
\subcaption{In-degree Centrality}
\end{minipage}%
    \caption{Robustness of the top nodes as the function of fraction of manipulated edges \textit{f}}
    \label{centralityrobust}
\end{figure}
The error level is considered as a specific percentage value from the set of {1\%, 5\%, 10\%, 20\%}, that is relative to the number of manipulated edges from the original true network. \autoref{centralityrobust} shows the results of the different Centrality measure as the function of the fraction of manipulated edges $f$. For each fraction level, the test is repeated for 100 times, and the graphs show the average of the all sampled sets. \autoref{pearson} shows the Pearson correlation between the nodes centrality in manipulated network and the original network when 10\% of the edges are randomly deleted.

\begin{table}[h]
\centering{}\protect\caption{Node centrality Pearson correlation between the manipulated network and the original network}
\label{pearson}
\adjustbox{max height=\dimexpr\textheight-6.5cm\relax,
           max width=\textwidth}{
\begin{tabular}{ l  l  l  l }
\hline
Measure & Katz prestige centrality & Betweenness centrality & Degree centrality\\ 
 \hline
 Significance (p-value) & $2.2e^-16$ & $0.001$ & $0.003$ \\ 
 Correlation & $0.939$  & $0.948$ & $0.67$ \\ 
 \hline	
\end{tabular}
}
\end{table}

The pattern and level of robustness of the three selected centrality measures considered in this paper are are not as similar as suggested in \cite{CRobust2}. In-degree centrality shows more fragility comparing to betweenness and Katz measures. This difference could be related to the network topology as suggested by \cite{CRobust}. The results also confirm the findings of \cite{CRobust} \cite{CRobust2} that accuracy declines monotonically with increasing error.

As can be seen in the graph, the Katz centrality is fairly robust to the edge deletion when less than 20 percent of the network structure is touched. The graphs indicate a moderate fragility when the network structure is hugely manipulated. For example the removal of 20 percent of the edges somehow impacts the in-degree centrality. However in more than 90 percent of these extreme samples, the top 1 node in the manipulated network is a member of top ten percent of nodes in the original graph. The results imply that centrality measures on Legislation Network are quite robust under small amounts of error (such as 5 percent or under) and to some extent fragile under bigger data errors. So the reliability of the network information is very important for in-depth network studies. As explained earlier the the precision and the recall of the proposed Information Extraction process is above 98 percent, so it is reasonable to compute the centrality measures when studying the Legislation Network.  

\section{Conclusion}
This study focused on the \textbf{time} as a very important attribute in understanding and analyzing legislation. Legislation Network has been discussed in recent years, but the importance of having access to the historic legislation was never discussed much. This paper underlined the value of studying legislation as dynamic networks, and proposed a new Information Extraction process to achieve a highly accurate Legislation Network. The performance of the data extraction framework is examined, is compared to the previous studies and proved to be considerably high. This work contributed to the literature of network Information Extraction from old documents, and insisted on the value and applications of the dynamic Legislation Network. The proposed process can be used not only in the legal domain but also in various research areas involving documented knowledge, facts, and cases. 

Analyzing a dynamic Legislation Network is a novel approach to understand the underlying process behind the generation of the laws, and to study the behaviour, culture and growth of societies. This subject is very interesting, but mathematically complicated. So it will be discussed in a separate study.

\end{document}